\documentclass[iop,apj]{emulateapj}
\usepackage{epsfig}
\usepackage{amsmath, amsthm, amssymb}
\usepackage[pdftex, plainpages=false, colorlinks=true, anchorcolor=blue, linkcolor=blue, citecolor=blue, bookmarks=false]{hyperref}
\usepackage{color}
\usepackage{times}
\usepackage{microtype}
\usepackage{float}

\newcommand{\beq}{\begin{equation}}
\newcommand{\eeq}{\end{equation}}
\newcommand{\bea}{\begin{eqnarray}}
\newcommand{\eea}{\end{eqnarray}}

\slugcomment{Submitted to the Astrophysical Journal}

\begin{document}

\title{The Dependence of the Neutrino Mechanism of Core-Collapse Supernovae on the Equation of State}
\author{Sean M. Couch\altaffilmark{1}}

\affil{Department of Astronomy \& Astrophysics, Flash Center for Computational Science, University of Chicago, Chicago, IL, 60637; smc@flash.uchicago.edu}
\altaffiltext{1}{Hubble Fellow}

\shorttitle{CCSNe NEUTRINO MECHANISM EOS DEPENDENCE}
\shortauthors{COUCH}

\begin{abstract}

We study the dependence of the delayed neutrino-heating mechanism for core-collapse supernovae on the equation of state.  Using a simplified treatment of the neutrino physics with a parameterized neutrino luminosity, we explore the relationship between explosion time, mass accretion rate, and neutrino luminosity for a 15 $M_\sun$ progenitor in 1D and 2D.  We test three different equations of state commonly used in core-collapse simulations: the models of \citet{Lattimer:1991fz} with incompressibility of 180 MeV and 220 MeV, and the model of \citet{Shen:1998kx}, in order of increasing stiffness.  We find that for a given neutrino luminosity the time after bounce until explosion increases with the stiffness of the equation of state:  the Lattimer \& Swesty EOS explode more easily than that of Shen et al.  We find this holds in both 1D and 2D, while for all models explosions are obtained more easily in 2D than in 1D.  We also discuss the relevance of approximate instability criteria to realistic simulations. 


\end{abstract}

\section{Introduction}
\label{sec:intro}

The exact process that halts the collapse of the core of a massive star at the end of its life and drives a successful supernova explosion, releasing a multitude of neutrinos and ejecta with around $10^{51}$ erg of kinetic energy is not yet fully-understood.  The delayed neutrino-heating mechanism of core-collapse supernovae \citep[CCSNe,][]{Colgate:1966cl,Bethe:1985da}, the leading candidate for the explosion mechanism, fails to consistently produce energetic explosions for the wide range of possible progenitor masses in simulations that include high-fidelity treatments of neutrino transport.  Neutrinos are an attractive explanation for the explosion mechanism as the bulk of the gravitational binding energy of the progenitor core, some $10^{53}$ erg, will be radiated away as neutrinos as the proto-neutron star (PNS) forms and cools; only about $10^{51}$ erg of this energy needs to be transferred to the collapsing stellar material in order to explain typical CCSNe energies.  Only a fraction, however, of the core's original gravitational binding energy, perhaps a few times $10^{52}$ erg, will be released as neutrino radiation in the first second following the collapse and bounce of the core, approximately the timescale on which a successful explosion must occur.  Neutrino radiation-hydrodynamic simulations of core collapse show that this is challenging to achieve, particularly in 1D simulations.  It is now clear that multi-dimensional effects, such as proto-neutron star convection \citep{Epstein:1979tg,Burrows:1993ki,Dessart:2006cg,Burrows:2007kha}, neutrino-driven convection \citep{1996A&A...306..167J}, the standing accretion shock instability \citep[SASI,][]{Blondin:2003ep}, and turbulence \citep{Murphy:2011ci} play a critical role in the success of the neutrino mechanism in driving explosions.  Together, these multidimensional effects can push some progenitors over the critical threshold, resulting in somewhat marginal explosions in 2D in certain simulations \citep{Marek:2007vi,Bruenn:2009cw,Suwa:2010wp,Mueller:2012tp}.  This result is dependent on details of the numerical scheme and treatment of neutrino transport; the 2D simulations of \citet{Burrows:2006js,Burrows:2007kh} do not find neutrino-driven explosions for any progenitors.\footnote{ \citet{Burrows:2006js,Burrows:2007kh} find explosions via the ``acoustic'' mechanism first revealed by their simulations.  The work of \citet{Weinberg:2008ky}, however, shows that the  $\ell=1$ g-mode vibration of the neutron star which powers the acoustic mechanism will in reality saturate at energies well below that required to drive an explosion.}  Based on a series of general relativistic 1D simulations including Boltzmann transport for neutrinos, \citet{Lentz:2012fy} suggest that the discrepancies in the results from the various groups simulating CCSNe may be due in large part to differences in how three major aspects of the physics are treated: general relativistic versus Newtonian dynamics, the neutrino interactions and opacities considered, particularly inclusion of inelastic neutrino scattering, and observer frame corrections in the neutrino transport.

\begin{deluxetable*}{ccccccc}
\tablecolumns{7}
\tabletypesize{\scriptsize}
\tablecaption{
EOS Parameters and experimental limits.
\label{table:eosParams}
}
\tablewidth{0pt}
\tablehead{
\colhead{EOS} &
\colhead{$K$ \tablenotemark{a}} &
\colhead{$K'$ \tablenotemark{b}} &
\colhead{$J$ \tablenotemark{c}} &
\colhead{$L$ \tablenotemark{d}} &
\colhead{$M^{\rm max}_{\rm NS}$\ \tablenotemark{e}} &
\colhead{$R_{1.4}$ \tablenotemark{f}} \\
\colhead{} &
\colhead{(MeV)} &
\colhead{(MeV)} &
\colhead{(MeV)} &
\colhead{(MeV)} &
\colhead{($M_\sun$)} &
\colhead{(km)} 
}

\startdata
LS180 & 180 & -451 & 28.6 & 74 & 1.84 & 12.2\\
LS220 & 220 & -411 & 28.6 & 74 & 2.06 & 12.7 \\
LS375 & 375 & -162 & 28.6 & 74 & 2.72  & 13.5 \\
STOS  & 281 & -285 & 36.9 & 111 & 2.22 & 14.6 \\
HS (TMA) & 318 & -572 & 30.7 & 90 & 2.02 & 13.5 \\
 & & & & \\
Limits & $240\pm10$ & -$355\pm95$ & $\sim 32$ & $\sim 75$ & $>1.97\pm0.04$ & 11 - 12
\enddata

\tablenotetext{a}{Nuclear incompressibility.}
\tablenotetext{b}{Skewness coefficient.}
\tablenotetext{c}{Symmetry energy coefficient.}
\tablenotetext{d}{Symmetry energy slope.}
\tablenotetext{e}{Maximum gravitational mass of a cold neutron star.}
\tablenotetext{f}{Radius of a 1.4 $M_\sun$ neutron star.}
\end{deluxetable*}

All of this indicates, unsurprisingly, that simulations of the neutrino mechanism are sensitive to microphysics, and details of how the microphysics is treated.  It is still uncertain, however, if the variation in the handling of microphysics is large enough to explain the dearth of energetic explosions; that is, if a more accurate handling of the details of the neutrino physics will result in successful explosions having around $10^{51}$ erg of kinetic energy for a wide range of progenitor masses.  \citet{Nordhaus:2010ct} have suggested a larger impact on the success of the neutrino mechanism results from fully 3D simulations. Using a simplified neutrino heating and cooling scheme rather than expensive neutrino transport, \citet{Nordhaus:2010ct} find that the driving neutrino luminosity necessary to obtain an explosion is about 20\% lower in 3D than in 2D.  This difference may be larger than what we could expect to gain from higher-fidelity treatments of neutrino transport or inclusion of general relativity in simulations.  \citet{Hanke:2011vc} sought to reproduce this result.  Employing a neutrino heating/cooling approach similar to \citet{Nordhaus:2010ct}, \citet{Hanke:2011vc} find the critical neutrino luminosity in 3D is not significantly lower than that of 2D.  This raises the questions of just how beneficial 3D simulations are to the success of the neutrino mechanism and what is the cause for the differences in the results of  \citet{Nordhaus:2010ct} and \citet{Hanke:2011vc}?  As 3D simulations become more feasible and available, we may be able to answer the question of the importance of 3D.  Recently, \citet{Takiwaki:2012ck} report an explosion for an 11.2 $M_\sun$ progenitor in a 3D simulation performed using the isotropic diffusion source approximation for neutrinos \citep[IDSA,][]{Liebendorfer:2009kw}.  Compared to 2D, Takiwaki et al. found that the convection below the gain region was more vigorous in 3D enhancing the neutrino luminosity beyond the 2D case.  The somewhat modest resolution they were able to afford, however, means that the issue of the importance of 3D simulations cannot be settled on the basis of their results alone.

\begin{figure}
\centering
\includegraphics[width=3.5in]{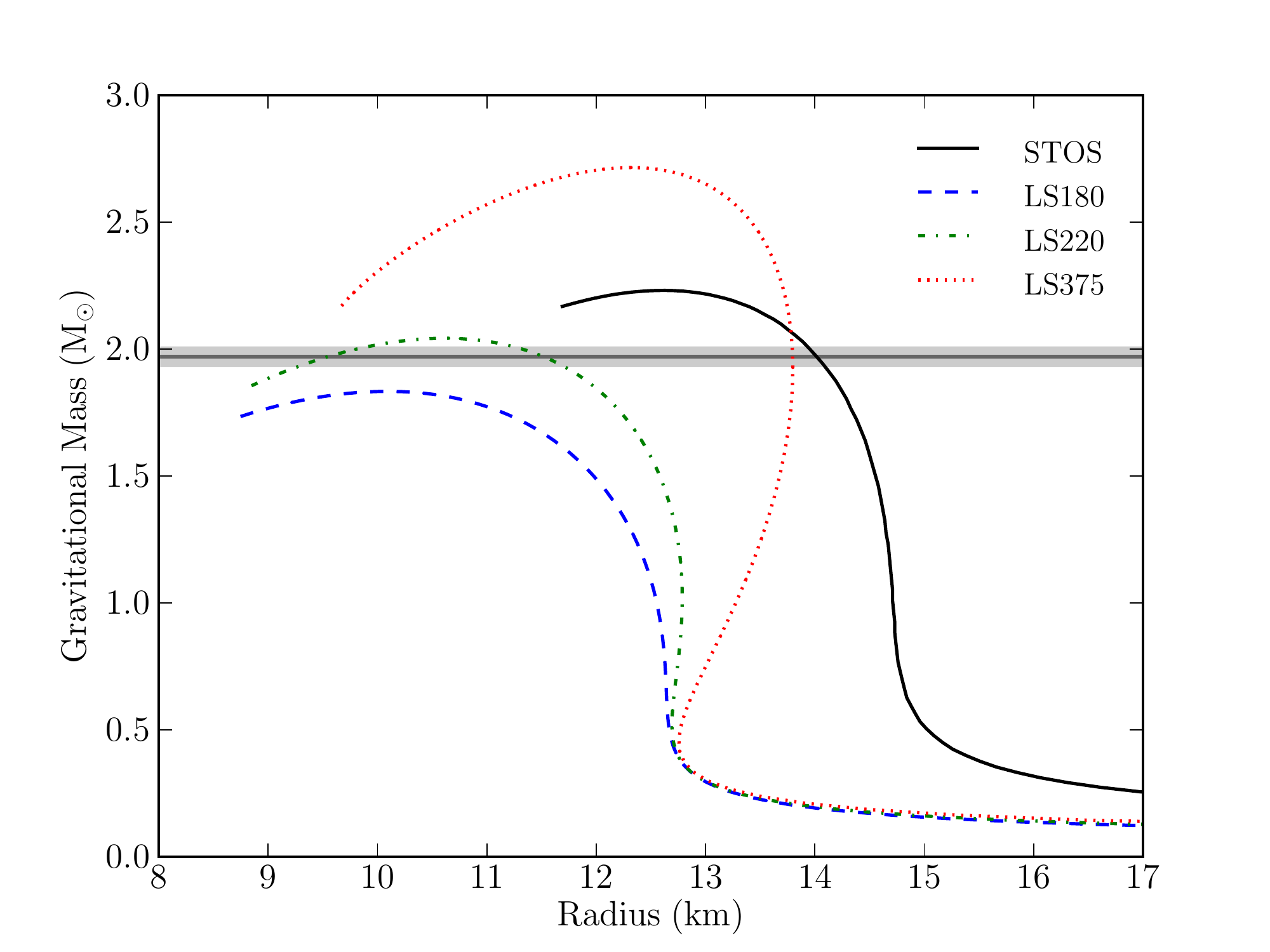}
\caption{Mass-Radius relations for cold neutron stars as computed by solving the Tolman-Oppenheimer-Volkov equations for the various EOS listed.}
\label{fig:TOV}
\end{figure}

Besides the neutrino physics, an important microphysical dependence of the CCSN mechanism is the equation of state.  The equation of state for matter at densities and temperatures relevant to CCSNe is still an active area of research \citep[e.g.,][]{Hempel:2010fh,Shen:2010wa,Shen:2011ta,Shen:2011wk,2011ApJS..197...20S,Furusawa:2011ck}.  Our understanding of  nuclear interactions at these densities and temperatures is incomplete and a number of models exist.  Testing the various EOS models with experiment and observation is challenging, but a few constraints do exist.  \citet{Hempel:2012bh} review some of the constraints on the EOS parameters from laboratory experiments.  In Table \ref{table:eosParams} we summarize the constraints on the nuclear EOS parameters and the respective numbers for five EOS models used in CCSN simulations: Lattimer \& Swesty with incompressibility parameters of 180 MeV (LS180), 220 MeV (LS220), and 375 MeV (LS375), Shen et al. (STOS), and the recent EOS of \citet[][]{Hempel:2010fh} using the TMA model for nuclear interactions (HS (TMA)).  Most values for the limits are taken from experiment \citep[for references to the experimental data see][]{Hempel:2012bh}, except for the skewness coefficient, $K'$, which is only a theoretical estimate. Based on the measurements listed in the table, only LS220 falls within or near the limits for each of the EOS parameters.

Observations of neutron stars and pulsars can also place constraints on the EOS for hot dense matter.  Using a parameterized EOS, \citet{Steiner:2010el} analyze a small sample of neutron stars and determine a most likely mass-radius relationship for cold neutron stars.  Their results favor a radius of $11-12$ km for a 1.4 $M_\sun$ neutron star, and a maximum neutron star mass in the range $1.9-2.2$ $M_\sun$.  The maximum mass of a cold neutron star is an important constraint for the EOS models.  The observations of PSR J1614-2230 by \citet{Demorest:2010bf} place a very tight limit on the mass of the neutron star in this system of $1.97\pm0.04\ M_\sun$.  Given an EOS model, solutions to the Tolman-Oppenheimer-Volkov equation yield the gravitational mass-radius relationship for cold neutron stars.  Figure \ref{fig:TOV} shows the mass-radius relationships for the EOS of \citet{Lattimer:1991fz} and \citet{Shen:1998kx}, along with the mass limits for PSR J1614-2230.  Based on the measurement of \citet{Demorest:2010bf}, LS180 is ruled out as it does not produce a large enough maximally-massed neutron star.  The Shen et al. EOS produces a maximum neutron star mass well above the limit placed by PSR J1614-2230 (2.22 $M_\sun$), but STOS predicts a radius for a 1.4 $M_\sun$ NS of around 15 km, well outside the limits estimated by \citet{Steiner:2010el}.  The theoretical predictions for the radius of a 1.4 $M_\sun$ NS of \citet{Hebeler:2010dx} are somewhat broader: 9.7 - 13.9 km, but STOS still lies outside of this range.  Only LS220 satisfies the constraint on the maximum NS mass from PSR J1614-2230 and produces a 1.4 $M_\sun$ NS with a radius  ($\sim 12.7$ km) close to the upper estimate of \citet{Steiner:2010el} and well within the limits of \citet{Hebeler:2010dx}.  Figure \ref{fig:TOV} also shows the mass-radius relationship for LS375.  This EOS easily satisfies the requirement on the maximum neutron star mass, but falls above the radius limits of \citet{Steiner:2010el} and has an incompressibility (375 MeV) well above that found in experiment.

In this work, we study the dependence of the neutrino mechanism on the equation of state employed.  Using 1D and 2D hydrodynamic simulations with simplified neutrino transport, we determine the explosion times for a 15 $M_\sun$ progenitor as a function of neutrino luminosity for three EOS: \citet{Lattimer:1991fz} with $K=180$ MeV and $K=220$ MeV, and \citet*[][]{Shen:1998kx}.  In 1D only, we also run simulations using LS375.  Our treatment of the neutrino physics is similar to those used by \citet{Murphy:2008ij}, \citet{Nordhaus:2010ct}, and \citet{Hanke:2011vc}: we assume local neutrino heating and cooling, based on the rates derived by \citet{Janka:2001fp}, and a constant neutrino luminosity. We find that the explosion times for a given neutrino luminosity are significantly dependent on the EOS used, with the general trend of easier explosions for softer equations of state.  Thus, the Lattimer \& Swesty models result in easier explosions than the Shen et al. EOS, with LS180 leading to the earliest explosions at a given neutrino luminosity.  Early 1D simulations using a simple, parameterized EOS showed that stronger shocks result for softer EOS \citep{Baron:1985ec, Baron:1985cd}.  Our results show that this trend is recovered for more realistic nuclear EOS and in 2D simulations.

The EOS has been shown to influence the results of CCSN simulations in a number of contexts. \citet{Thompson:2003kn} explored the sensitive of their 1D neutrino radiation hydrodynamic simulations to the value of $K$ for the Lattimer \& Swesty EOS.  They found that the overall evolution during the first 200 ms post-bounce is not very dependent on $K$: the resulting temperatures were slightly higher and the emergent neutrino luminosities at most 9\% larger for LS180 as compared to LS375.  \citet{Sumiyoshi:2005kg} compared LS180 and STOS in 1D general relativistic simulations.  While neither model explodes in their simulations, they find differences in composition resulting from the different symmetry energies (see $J$ in Table \ref{table:eosParams}) and in temperature evolution with LS180 giving slightly larger temperatures and, therefore, slightly higher neutrino luminosities.  The higher neutrino luminosity and heating rate found for LS180 is somewhat counteracted by the more compact proto-neutron star that results from the lower incompressibility.  \citet{Sumiyoshi:2005kg} find that the differences between the two models grow with time, particularly after 300 ms post-bounce.  \citet{Marek:2009bx} explored the dependence on the gravitational wave signal produced by 2D simulations on the EOS, comparing LS180 and the stiffer EOS of \citet{Hillebrandt:1985to}.  They find that the matter-generated gravitational wave signal is not very dependent on the EOS but that the differences in the anisotropic neutrino emission gravitational wave signal, which can dominate the matter gravitational wave signal, are significantly different between LS180 and Hillebrandt \& Wolff EOS.  In failed explosions, the formation of black holes has been found to be highly dependent on the EOS \citep{Sumiyoshi:2007cd, Fischer:2009ka, OConnor:2011hk, Hempel:2012bh}.  Generally, stiffer EOS result in a greater time to black hole formation following bounce.

This paper is organized as follows.  In Section \ref{sec:Method} we describe the details of our numerical approach.  In Section \ref{sec:Results} we describe the results of our simulations and the dependence of the explosions on the EOS.  In Section \ref{sec:Conclusions} we conclude and discuss how our work relates to recent work on neutrino-driven supernovae.

\section{Computational Methodology}
\label{sec:Method}

\begin{figure}
\centering
\includegraphics[width=3.5in]{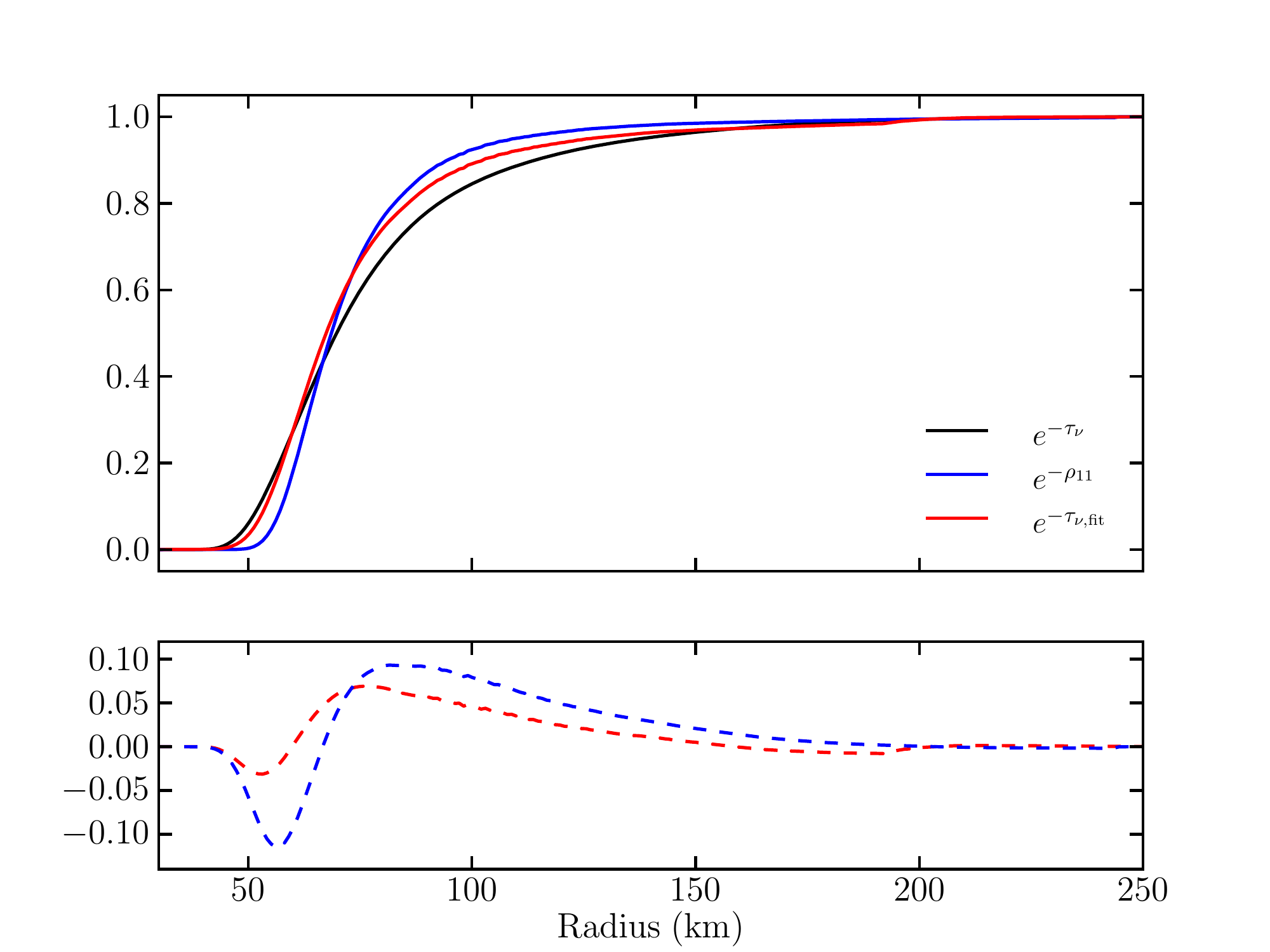}
\caption{Comparison of the fitting formula used in this work to compute the neutrino optical depth and the integrated optical depth from a one-dimensional simulation at 200 ms post-bounce.  The black curve is the exponential of the negative neutrino optical depth as computed using the effective of opacity of \citet{Janka:2001fp}. The red curve is the same but using the fitting formula.  Plotted also is an exponential cutoff based on the density in units of $10^{11}$ g cm$^{-3}$, $\rho_{11}$.  The bottom panel shows the error between the approximate approaches and that based on the computed optical depth.  The fitting formula used in this work matches the integrated optical depth well, better than the simple density-based approach.}
\label{fig:tauNu}
\end{figure}

We conduct our study using the FLASH\footnote{Available for download at www.flash.uchicago.edu.} simulation framework \citep{Dubey:2009hh} to solve the Eulerian equations of hydrodynamics,
\begin{align}
\frac{\partial \rho}{\partial t} + \pmb{\nabla} \cdot (\rho \pmb{ v}) &= 0,\label{eq:massCons} \\ 
\frac{\partial \rho {\pmb v}}{\partial t} + \pmb{\nabla} \cdot (\rho \pmb{ v v}) + \pmb{\nabla} P &= -\rho \pmb{\nabla} \Phi, \\
\frac{\partial \rho E}{\partial t} + \pmb{\nabla} \cdot [(\rho E + P){\pmb v}] &= \rho {\pmb v} \cdot \pmb{\nabla} \Phi + \rho (\mathcal{H-C}), \label{eq:enerCons}
\end{align}
where $\rho$ is the mass density, $\pmb v$ the velocity vector, $P$ the pressure, $\Phi$ the gravitational potential, $E$ the total specific energy, and $(\mathcal{H-C})$ is the difference between neutrino heating and cooling.  We use the directionally-unsplit hydrodynamics solver in FLASH to solve equations (\ref{eq:massCons}) - (\ref{eq:enerCons}) \citep[Lee \& Couch, in prep, see also][]{Lee:2009kq}.  The unsplit solver in FLASH has many options for spatial reconstruction scheme, Riemann solver, slope limiter, etc.  For this work we use third-order piecewise-parabolic spatial reconstruction \citep[PPM,][]{1984JCoPh..54..174C}, a ``hybrid'' slope limiter, and the HLLC Riemann solver.  Inside of shocks, we use the HLLE Riemann solver in order to avoid the odd-even decoupling instability in 2D \citep{Quirk:1994co}.  This approach allows us to avoid the smearing of contact discontinuities that the HLLE solver is prone to while benefiting from the HLLE solver's immunity to odd-even decoupling at shocks.  The ``hybrid'' slope limiter \citep{Balsara:2004cz} applies the monotonized central ({\tt mc}) limiter to linear wave families (such as density) and the robust, slightly more diffusive {\tt minmod} limiter to non-linear, self-steepening wave families.   The gravitational potential, $\Phi$, is calculated using the new multipole Poisson solver available in FLASH4 (see section 8.10.2.2 of the FLASH4 User's Guide), however, we set $\ell_{\rm max} = 0$ yielding monopole gravity.

Following the approaches of \citet{Murphy:2008ij}, \citet{Nordhaus:2010ct}, and \citet{Hanke:2011vc} we use the neutrino heating and cooling rates derived by \citet{Janka:2001fp},
\begin{align}
\mathcal{H} = 1.544\times10^{20} \left (\frac{L_{\nu_e}}{10^{52}\ {\rm erg\ s^{-1}}} \right ) \left (\frac{T_{\nu_e}}{4\ {\rm MeV}} \right )^2 \nonumber \\
\times \left (\frac{100\ \rm km}{r} \right )^2 (Y_p + Y_n) e^{-\tau_{\nu_e}}\ \left [ \rm{\frac{erg}{g \cdot s}} \right ],\label{eq:heat}
\end{align}
and
\beq
\mathcal{C} = 1.399\times10^{20} \left (\frac{T}{2\ \rm MeV} \right )^6 (Y_p + Y_n) e^{-\tau_{\nu_e}}\ \left [ \rm{ \frac{erg}{g\cdot s}} \right ],\label{eq:cool} 
\eeq
where $L_{\nu_e}$ is the electron neutrino luminosity (note it is assumed that $L_{\bar \nu_e}=L_{\nu_e}$), $T_{\nu_e}$ is the electron neutrino temperature, $r$ is the spherical radius, $(Y_p + Y_n)$ is the sum of the neutron and proton number fractions, $\tau_{\nu_e}$ is the electron neutrino optical depth, and $T$ is the matter temperature.  In all of our calculations, $T_{\nu_e}$ is set to 4 MeV.  The heating term, equation (\ref{eq:heat}), is applied only after core bounce has occurred in our simulations.  The factor $e^{-\tau_{\nu_e}}$ is included to provide an exponential cutoff of the heating and cooling at high density where the material is optically-thick to neutrinos.  As such, we do not calculate the optical depth self-consistently but instead rely on a piece-wise fit of the optical depth as a function of density.  In Figure \ref{fig:tauNu} we compare our piece-wise fitting approach for $\tau_{\nu_e}$ with the integrated electron neutrino optical depth computed using the effective opacity given by \citet{Janka:2001fp}, $\kappa_{\rm eff} = 1.5\times10^{-7} \rho_{10} (k T_{\nu_e}/4\ {\rm MeV})^2$.  As shown, the piece-wise fit closely approximates the actual integrated optical depth.  We also show in Figure \ref{fig:tauNu} an exponential cutoff based directly on the density.  We see that our piecewise approach better approximates the integrated optical depth than this simple density based approach.  We stress that the factor $e^{-\tau_{\nu_e}}$ in Equations (\ref{eq:heat}) and (\ref{eq:cool}) is used simply as a cutoff for the heating and cooling functions at high-density and the exact values of the cutoff are less important than consistent application and use in all models considered in a study such as the present.

\begin{figure}
\centering
\includegraphics[width=3.5in]{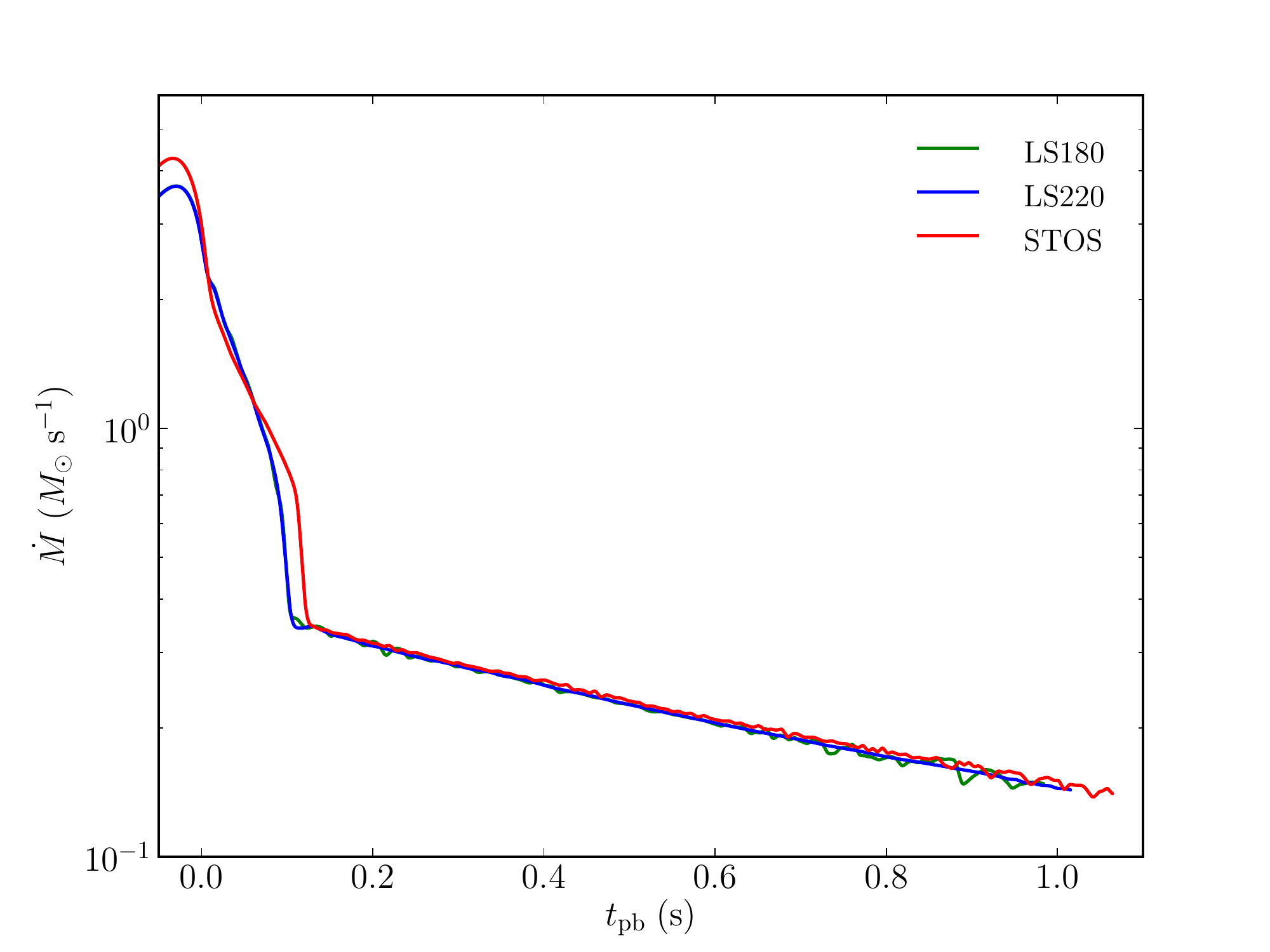}
\caption{Mass accretion rate as a function of time relative to bounce for the 15 $M_\sun$ progenitor used in this study.  The mass accretion rate is measured at a radius of 500 km for a non-exploding model.  The accretion rates for the three EOS used are shown.  The Lattimer \& Swesty EOS differ from that of Shen et al. only at early times and are essentially identical after 120 ms post-bounce.}
\label{fig:massAcc}
\end{figure}

\begin{deluxetable}{ccccccc}
\tablecolumns{7}
\tabletypesize{\scriptsize}
\tablecaption{
Time post-bounce and accretion rates at time of explosion for the 15 $M_\odot$ progenitor with 0.5 km resolution.
\label{table:results}
}
\tablewidth{0pt}
\tablehead{
\colhead{} &
\multicolumn{2}{c}{LS180} &
\multicolumn{2}{c}{LS220} &
\multicolumn{2}{c}{STOS} \\
\colhead{$L_{\nu_e}$\tablenotemark{a}} &
\colhead{$t_{\textrm{exp}}$\tablenotemark{b}} &
\colhead{$\dot{M}_{\textrm{exp}}$\tablenotemark{c}} &
\colhead{$t_{\textrm{exp}}$} &
\colhead{$\dot{M}_{\textrm{exp}}$} &
\colhead{$t_{\textrm{exp}}$} &
\colhead{$\dot{M}_{\textrm{exp}}$} \\
\colhead{($10^{52}$ erg/s)} &
\colhead{(ms)} &
\colhead{($M_{\odot}$/s)} &
\colhead{(ms)} &
\colhead{($M_{\odot}$/s)} &
\colhead{(ms)} &
\colhead{($M_{\odot}$/s)} }
\startdata 
\cutinhead{1D} 
2.0 &  \nodata   &  \nodata  &  \nodata  &  \nodata    &  \nodata   &  \nodata   \\
2.1 &  \nodata   &  \nodata  &  \nodata  &  \nodata    &           &            \\
2.2 &  689       &  0.192    &  816      &  0.169      &  \nodata   &  \nodata   \\
2.3 &  374       &  0.254    &  571      &  0.211     &  943       &  0.153     \\
2.4 &  213       &  0.289    &  402    &  0.251     &  829       &  0.170      \\
2.5 &  193       &  0.312    &  196         &  0.311     &  380       &  0.262      \\
2.7 &  175       &  0.317    &  178       &  0.320     &  216       &  0.310      \\
2.9 &                &           &           &            &  200       &  0.314      \\
\cutinhead{2D}
1.3 &  813       &  0.182    &  \nodata   &  \nodata      & \nodata    & \nodata    \\
1.5 &  277       &  0.279    &  368       &  0.262        & 732        &  0.192    \\
1.7 &  203       &   0.307   &  246        &  0.295       & 374        &  0.263  \\
1.8 &          &             &  194        &  0.313       &          &     \\
1.9 &  178       &  0.315    &                 &                    & 249        & 0.298    \\
2.0 &  179       &  	0.318    &  171         &  0.322       & 217        & 0.313    \\
2.1 &         &                       &  171         &  0.322       & 201        & 0.317    \\
2.2 &         &                       &  167         &  0.324       & 193        & 0.322  \\
2.4 &         &                       &                  &                   & 183        & 0.322  \\
2.6 &         &                       &                  &                   & 168        & 0.328  \\
\enddata
\tablenotetext{a}{Electron-neutrino luminosity.}
\tablenotetext{b}{Time after bounce of onset of explosion. 
A ``...'' symbol indicates 
that the model does not explode during the simulated period of evolution.}
\tablenotetext{c}{Mass accretion rate at onset of explosion.}
\end{deluxetable}

\begin{figure*} [!htb]
\centering
\begin{tabular}{cc}
\includegraphics[width=3.25in,trim= 0.3in 0in 0.5in 0.5in,clip]{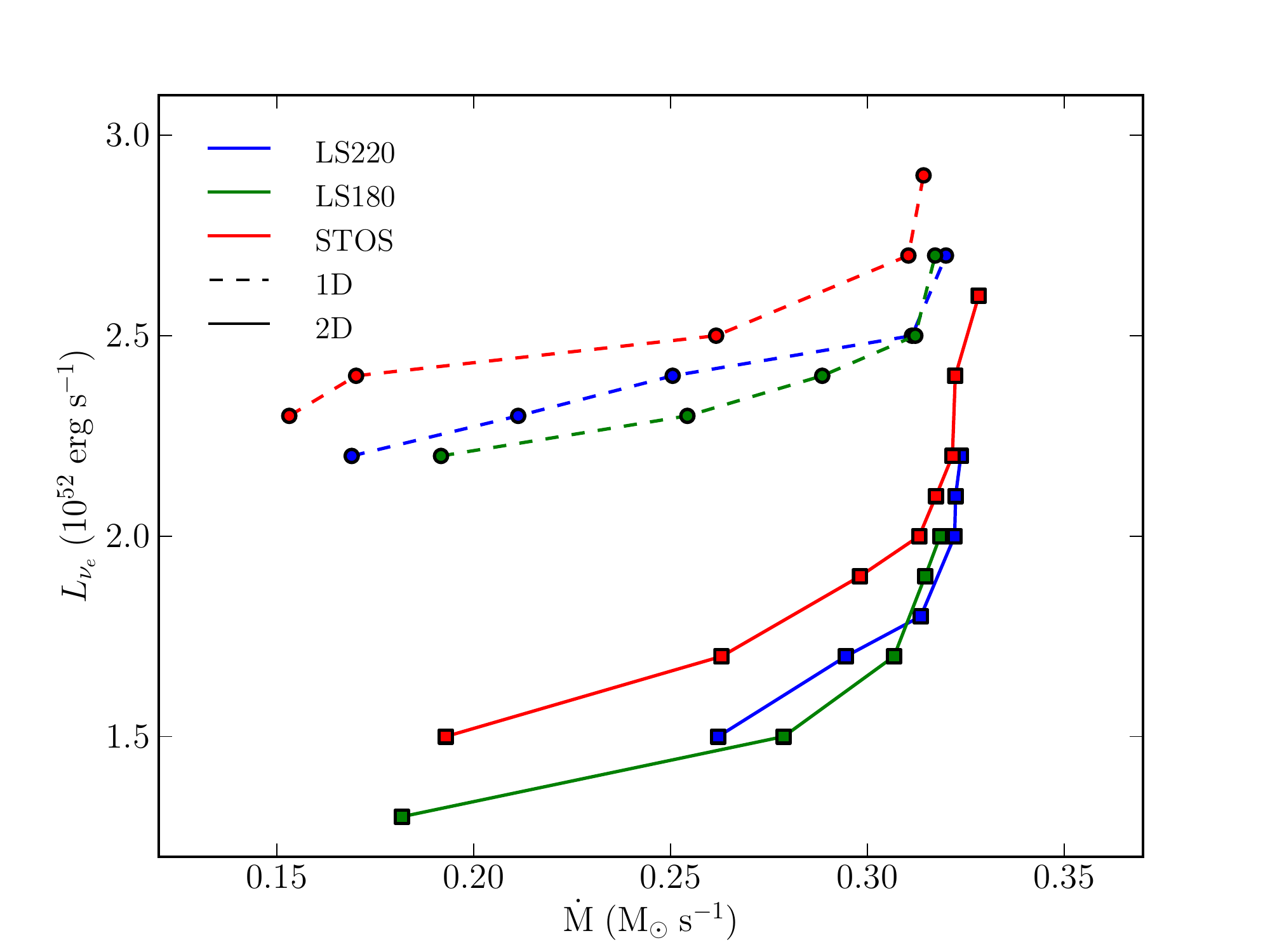} &
\includegraphics[width=3.25in,trim= 0.3in 0in 0.5in 0.5in,clip]{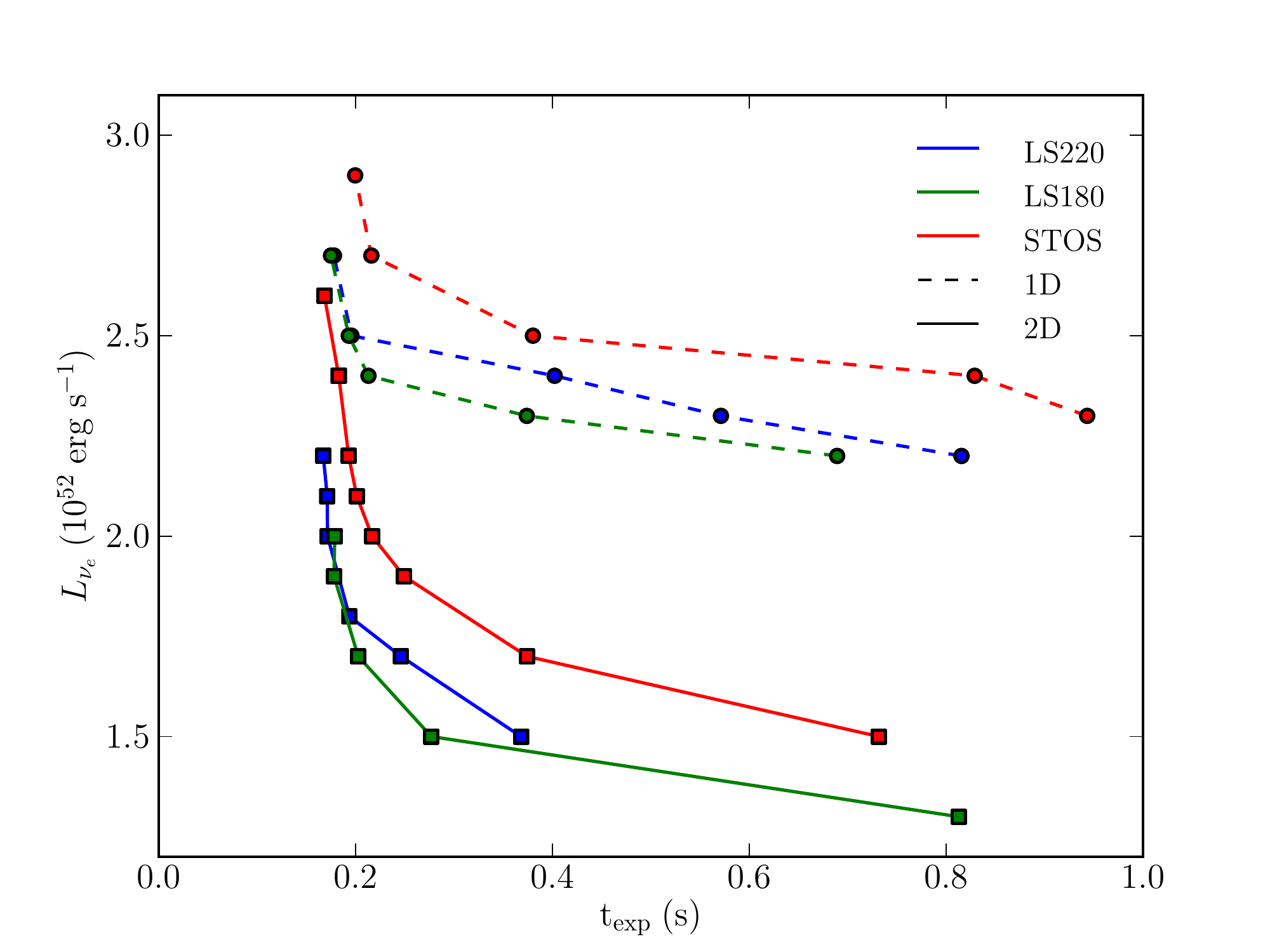}
\end{tabular}
\caption{Neutrino luminosity versus mass accretion rate at time of explosion (left panel) and time of explosion (right panel) for the three EOS considered in this work.  Results from both 1D (dashed lines) and 2D (solid lines) are shown. For each EOS explosions are found more easily in 2D than in 1D.  The Lattimer \& Swesty EOS result in easier explosions in both 1D and 2D than the Shen et al. EOS. }
\label{fig:mdot}
\end{figure*}

\begin{figure*}
\centering
\begin{tabular}{cc}
\includegraphics[width=3.25in,trim= 0.2in 0in 0.5in 0.5in,clip]{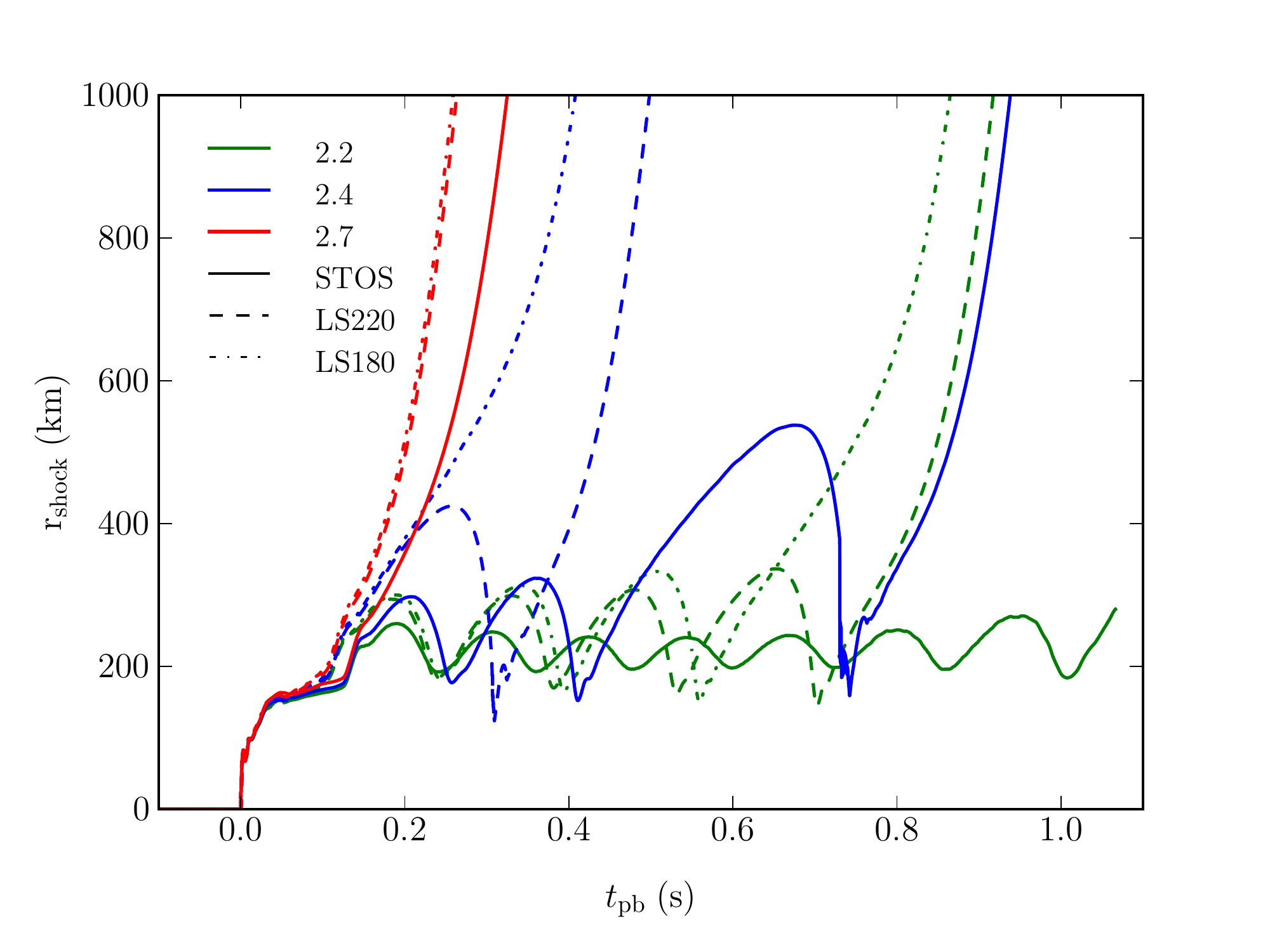} &
\includegraphics[width=3.25in,trim= 0.2in 0in 0.5in 0.5in,clip]{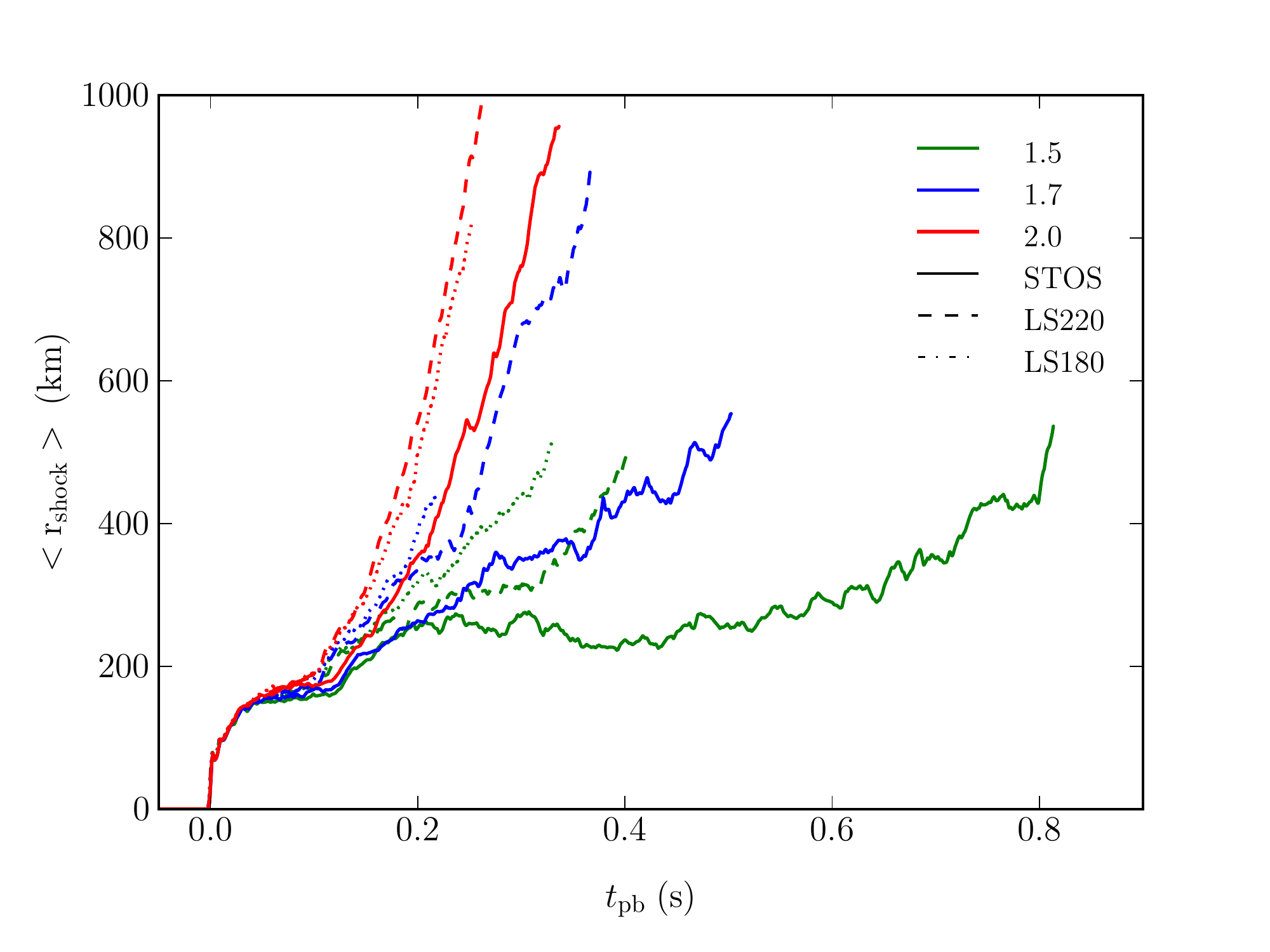}
\end{tabular}
\caption{Shock radii as a function of time for STOS, LS220, and LS180 in 1D (left) and 2D (right).  Three different neutrino luminosities are plotted for each EOS in each panel, as labeled.  }
\label{fig:rshock}
\end{figure*}

We calculate the evolution of the electron fraction, $Y_e$, according to the parameterization of \citet{Liebendorfer:2005ft} in which $Y_e$ is dependent only on density.  This approach is appropriate only for the pre-bounce collapse phase, however we employ the parameterization post-bounce as well in order to maintain consistency with \citet{Nordhaus:2010ct}. With no post-bounce evolution of the electron fraction the cooling above the neutrinosphere is grossly underestimated leading to significantly earlier and easier explosions, as mentioned by \citet{Hanke:2011vc}.  We do not include the entropy increases due to neutrino thermalization as prescribed by \citet{Liebendorfer:2005ft}. 

We run our simulations in 1D spherical and 2D cylindrical geometries using adaptive mesh refinement (AMR) as implemented in FLASH via PARAMESH \citep[v.4-dev,][]{2000CoPhC.126..330M}.  For this work we have extended FLASH's unsplit solver to spherical and cylindrical coordinates using an approach similar to those described in \citet{Skinner:2010vl} and Appendix A of \citet{Mignone:2005hc}.   Our fiducial set of simulations has a minimum grid-spacing of 0.5 km in each direction, obtained using 8 AMR levels.  The grid in our 1D spherical simulations covers radii from 0 km to 5000 km.  In 2D cylindrical runs, we simulate a grid that spans 5000 km in $R$ and 10,000 km in $z$.  We do not remove the proto-neutron star from our grid and follow the evolution from progenitor phase through collapse and bounce and into the shock stagnation/revival phase.  For boundary conditions at the outer edge of the domain, we assume radial power-law profiles that approximate the stellar envelope.  We find that simple ``outflow'' boundary conditions, which enforce a zero-gradient condition, are not appropriate for the core-collapse context because the amount of stellar material entering the outer edge of the domain will be overestimated, artificially enhancing the mass accretion rate at late times.  By experimentation we have found that ``outflow'' boundary conditions can, in this way, suppress explosions for neutrino luminosities near critical.  For all of our simulations, we use the 15 $M_\sun$ progenitor of \citet{Woosley:1995jn}.

We have incorporated the finite temperature equation of state routines of \citet{OConnor:2010bi} into the FLASH framework\footnote{These routines are available for download at stellarcollapse.org.}.  We use three different EOS models in our simulations, the models of \citet{Lattimer:1991fz} with incompressibility, $K$, of 180 MeV and 220 MeV and that of \citet{Shen:1998kx}.  Some parameters for these EOS are listed in Table \ref{table:eosParams}.

\section{Results}
\label{sec:Results}
\subsection{Dependence on EOS}

\begin{figure*}
\center
\begin{tabular}{cc}
\includegraphics[width=3.25in,trim= 0.35in 2.5in .6in 2in,clip]{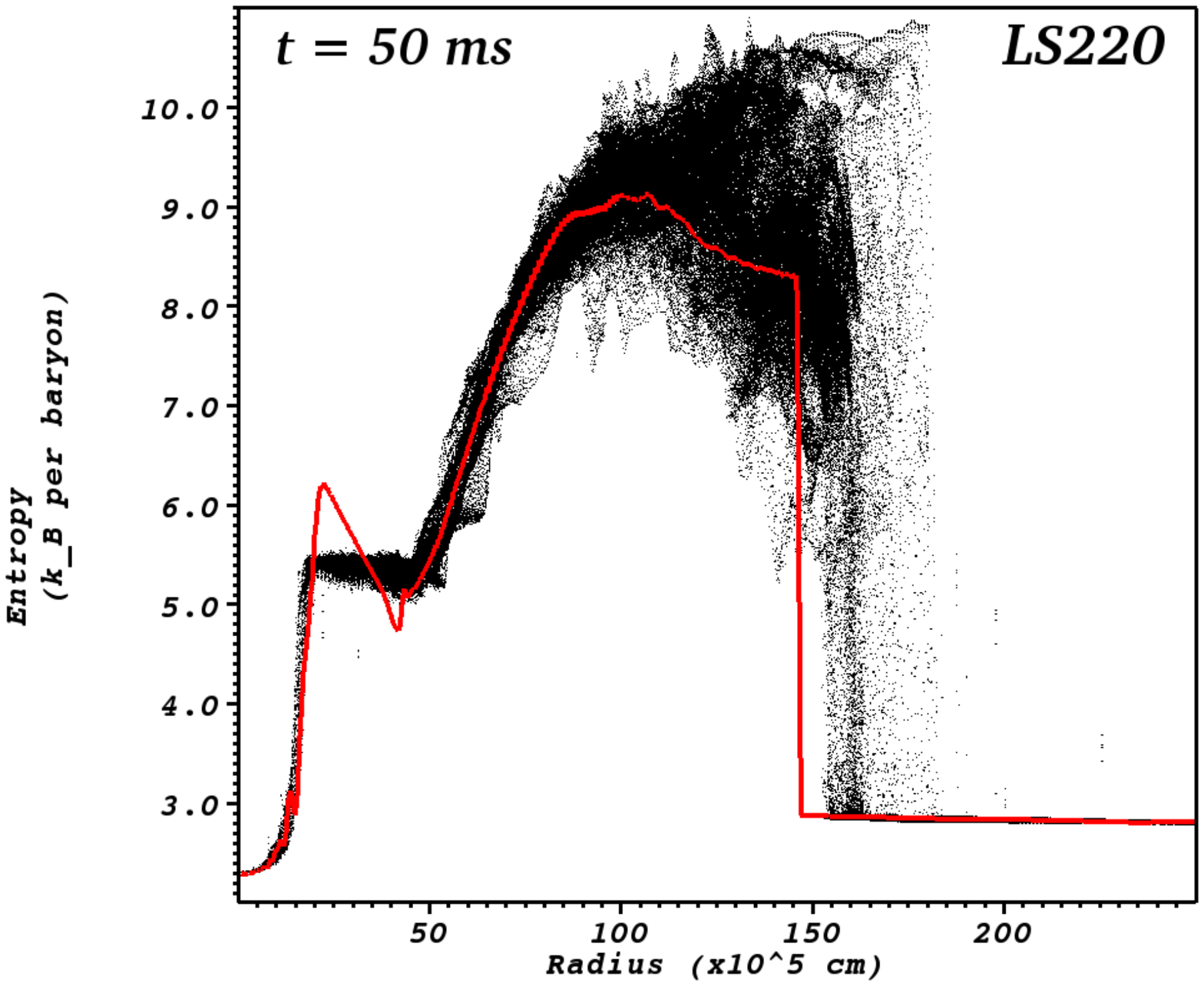} &
\includegraphics[width=3.25in,trim= 0.35in 2.5in .6in 2in,clip]{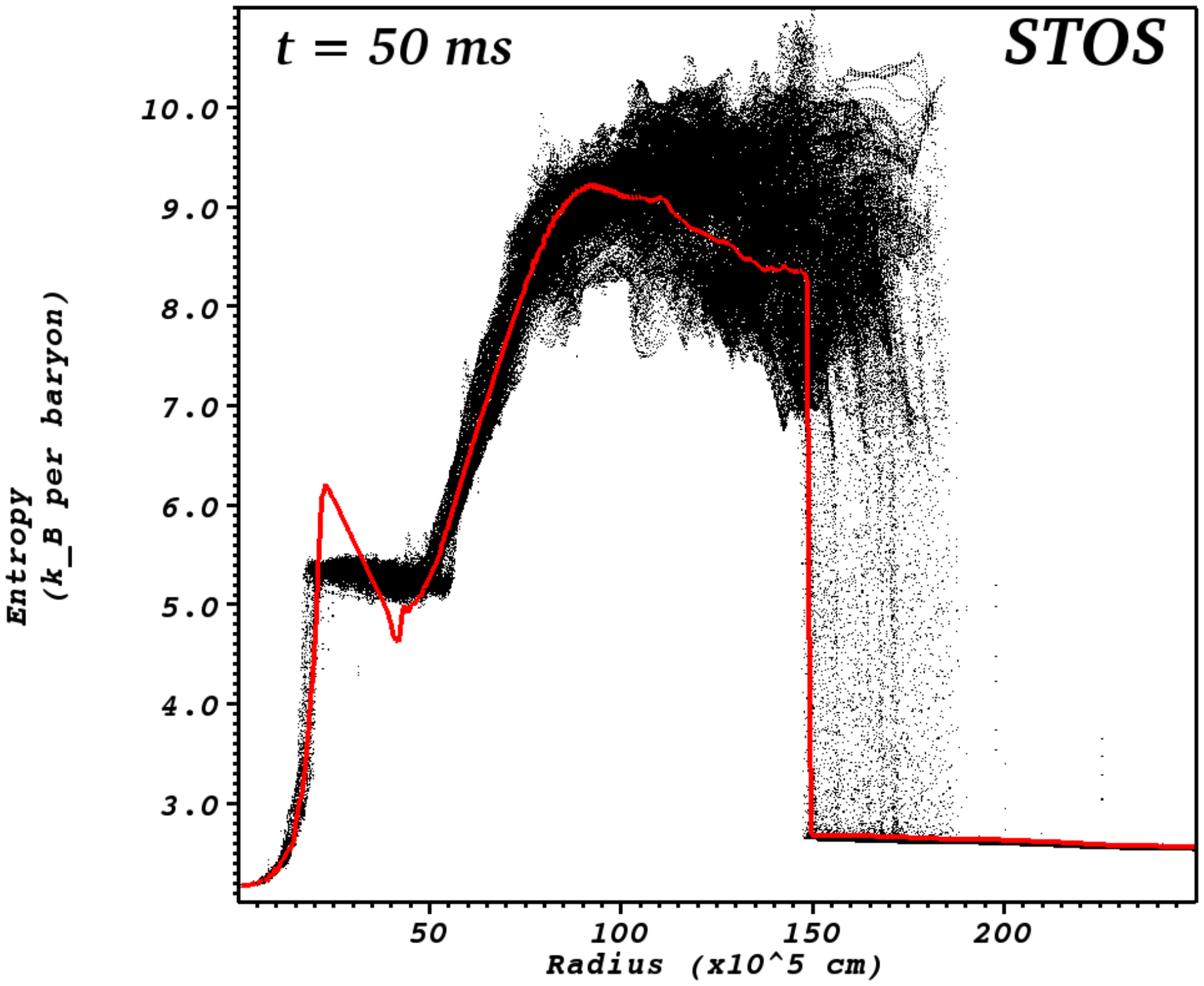}  \\
\includegraphics[width=3.25in,trim= 0.35in 2.5in .6in 2in,clip]{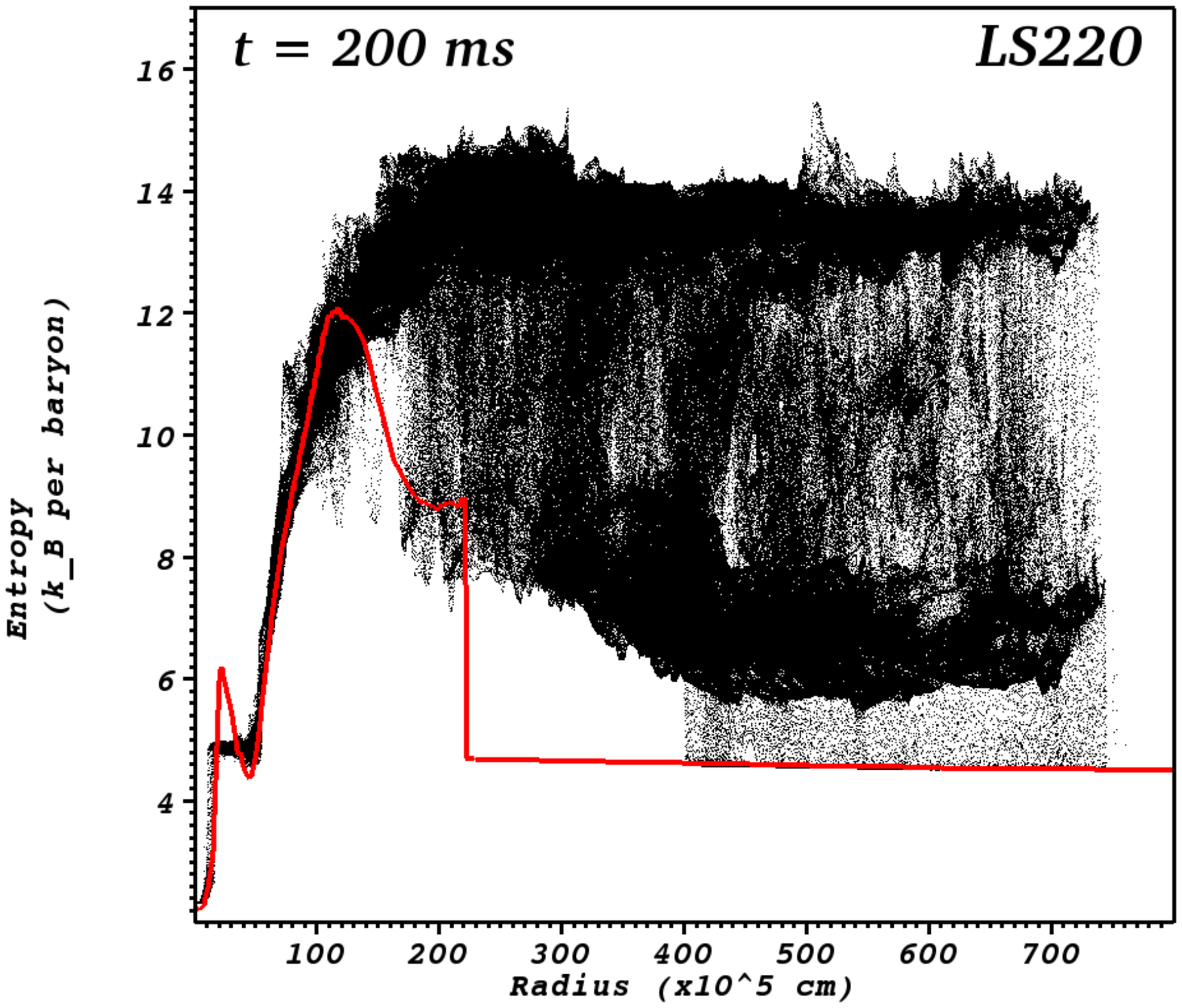} &
\includegraphics[width=3.25in,trim= 0.35in 2.5in .6in 2in,clip]{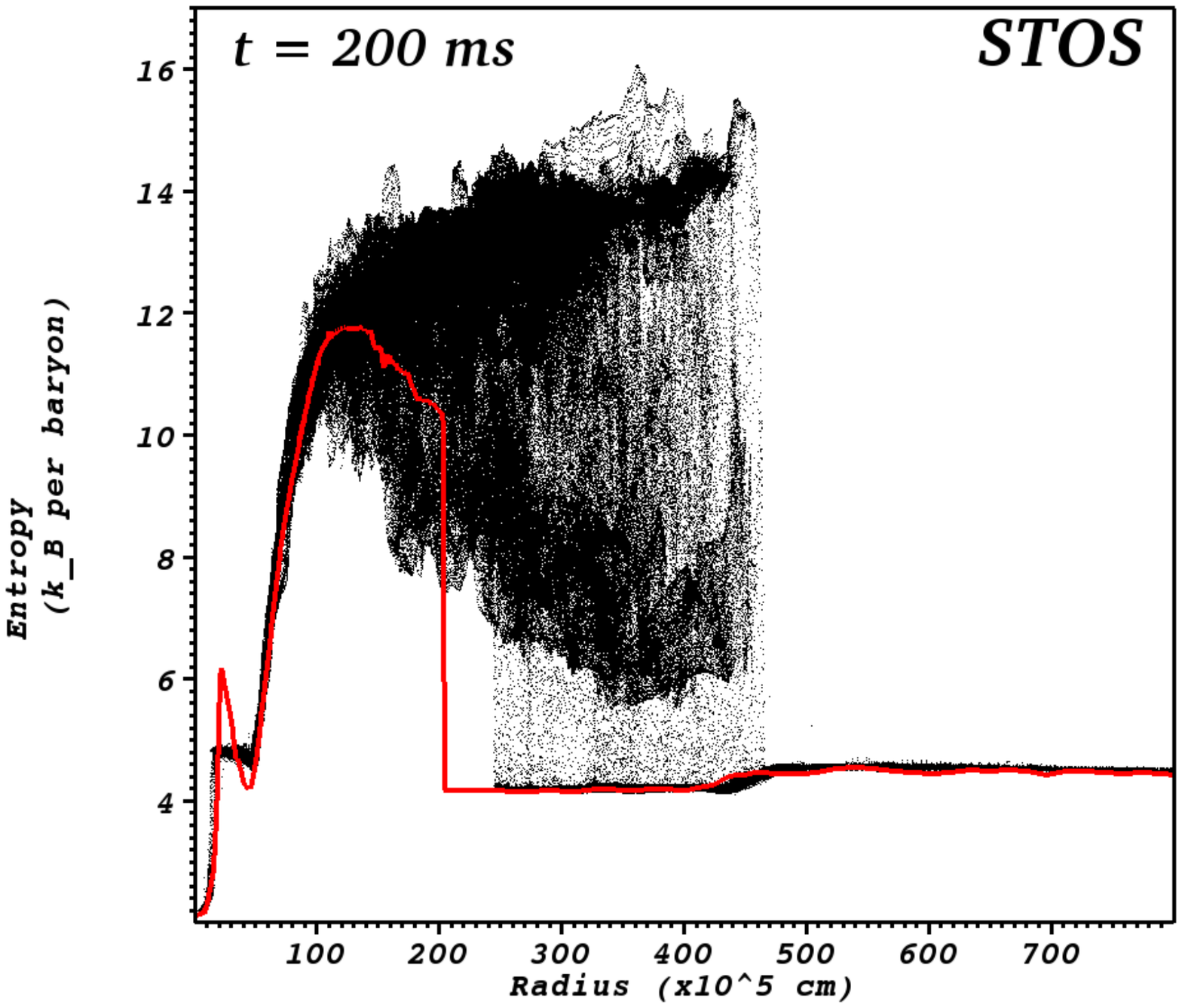}
\end{tabular}
\caption{Radial entropy plots comparing 1D (red lines) and 2D (black dots) for LS220 (left panels) and STOS (right panels) at two different times post-bounce:  50 ms (top) and 200 ms (bottom).  The neutrino luminosity for all the models shown in this figure is $2.0\times10^{52}\ {\rm erg\ s^{-1}}$. }
\label{fig:entr1d2d}
\end{figure*}

We ran a series of core collapse simulations in 1D and 2D in which we varied the driving neutrino luminosity.  For models that explode, we measure the post-bounce time of the explosion and the mass accretion rate at the time of explosion. We consider a model to have exploded if the average shock radius exceeds 400 km and does not subsequently fall back below 400 km. We measure the mass accretion rate at a radius of 500 km in both 1D and 2D simulations, though the measured mass accretion rates are identical since the 2D solution remains spherically-symmetric outside of the shock. Figure \ref{fig:massAcc} shows our measured mass accretion rates as a function of time post-bounce for the three EOS we consider. Our mass accretion rate is very similar to that of \citet{Hanke:2011vc} for their 15 $M_{\sun}$ progenitor model.  The differences in mass accretion history between STOS and LS can be attributed to models using STOS collapsing and bouncing a little faster than models using LS; bounce occurs 50 ms earlier for STOS as compared to LS.  Table \ref{table:results} summarizes the simulations we ran and the resulting explosion times and mass accretion rates at the time of explosion.  In Figure \ref{fig:mdot} we plot the driving neutrino luminosities as a function of explosion time and mass accretion rate at the time of explosion.

Our results show that the Lattimer \& Swesty EOS explode more easily than that of Shen et al., with LS180 resulting in the earliest explosions for a given neutrino luminosity (lowest curves in Fig. \ref{fig:mdot}).  The results then follow a basic trend that the stiffer the EOS, the harder it is to drive an explosion (LS180 being the softest EOS and STOS being the stiffest EOS we consider).  This trend holds in both 1D and 2D simulations.  For high neutrino luminosities, resulting in earlier explosions, the curves in Figure \ref{fig:mdot} for LS180 and LS220 converge, showing very little difference in time of explosion at these luminosities. Figure \ref{fig:rshock} shows the (average) shock radii in 1D (2D) for the three EOS and several neutrino luminosities.  The 1D simulations show a clear delineation of models into the three categories of the radial shock instability \citep{Fernandez:2012kg}:  oscillatory/stable\footnote{In our simulations, since the neutrino luminosity remains constant while the mass accretion rate decreases with time, there is really no such thing as a truly stable oscillatory model.  All models at a given neutrino luminosity will eventually explode once the mass accretion rate has dropped sufficiently.}, oscillatory/unstable, non-oscillatory/unstable.  In 2D, $\ell>0$ modes of the SASI dominate the purely radial ($\ell=0$) shock oscillations seen in 1D leading to a more chaotic variation in the average shock radius in time.  In Figure \ref{fig:entr1d2d} we compare the radial entropy profiles in 1D and 2D for LS220 and STOS at 50 ms and 200 ms post-bounce for $L_{\nu_e, 52} = L_{\nu_e}/(10^{52}\ {\rm erg\ s^{-1}}) = 2.0$.  For this luminosity, both 2D simulations result in rather prompt explosions whereas the 1D simulations fail to explode entirely with the shock stalling around 200 km.  The entropy scatter plots for the 2D simulations shown in Figure \ref{fig:entr1d2d} also demonstrate the entropy re-distribution due to convection and turbulence.

\begin{figure*}
\centering
\begin{tabular}{ccc}
\includegraphics[width=2.5in,trim= 0.4in 0.7in 3.4in 0in,clip]{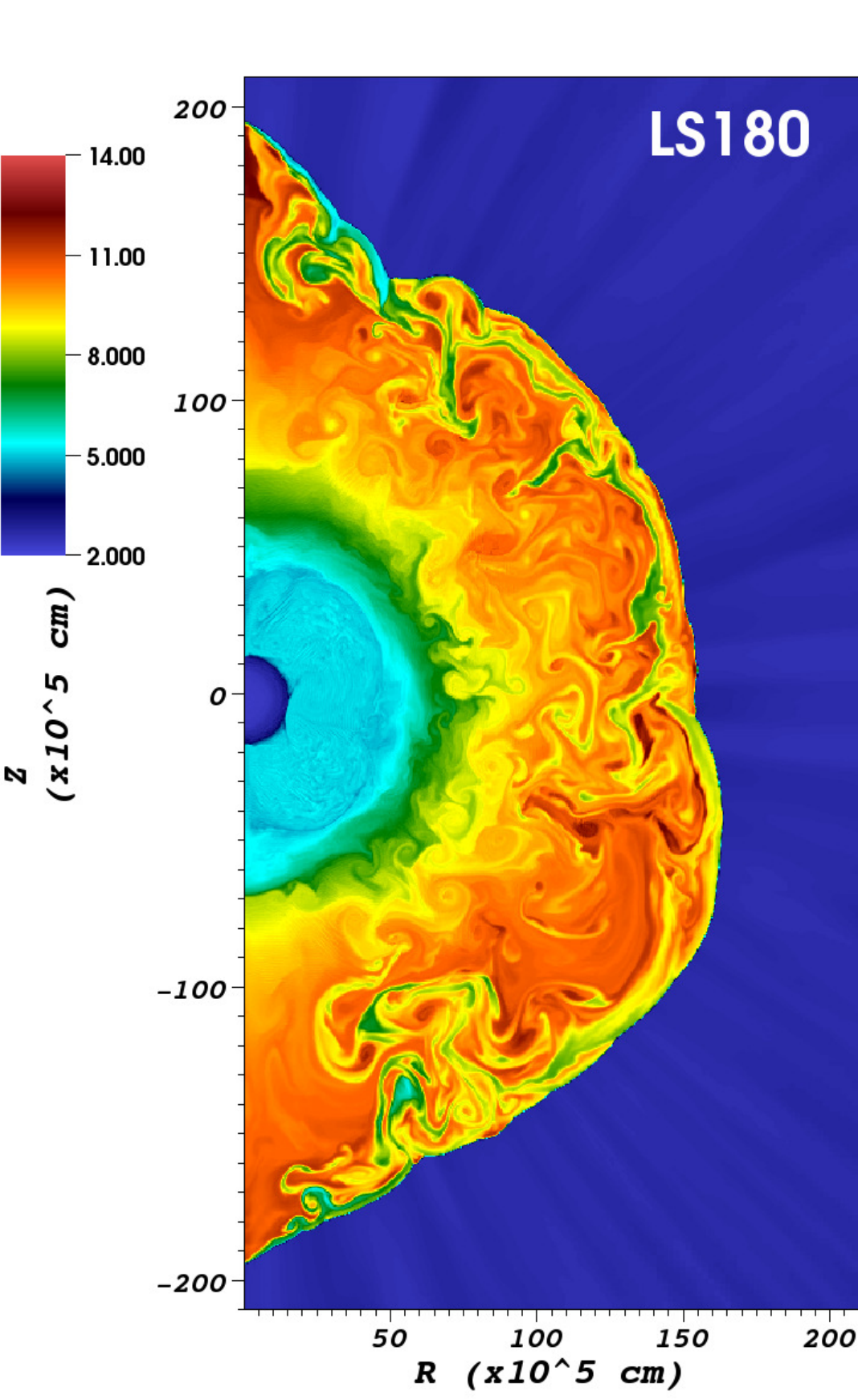}
\includegraphics[width=1.8in,trim= 1.6in 0.7in 3.4in 0.in,clip]{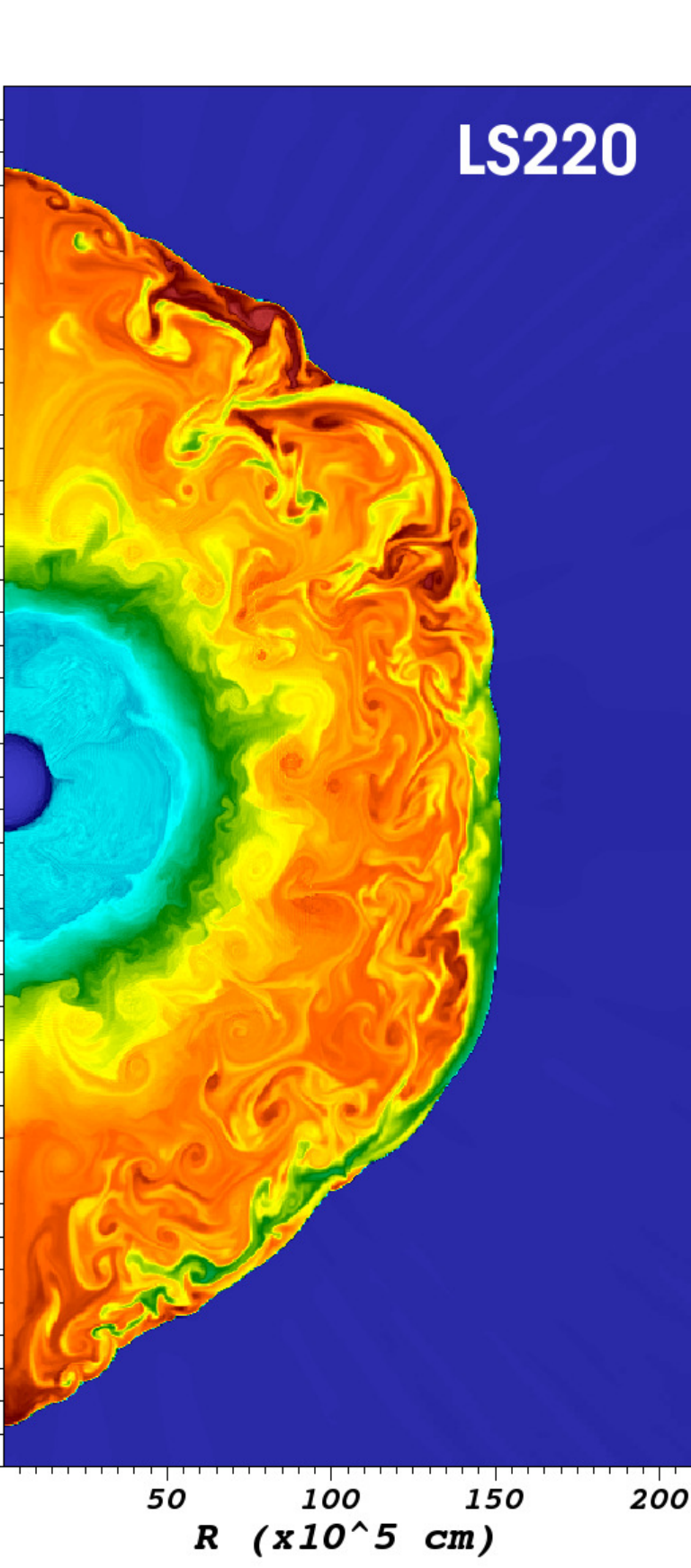}
\includegraphics[width=1.8in,trim= 1.6in 0.7in 3.4in 0in,clip]{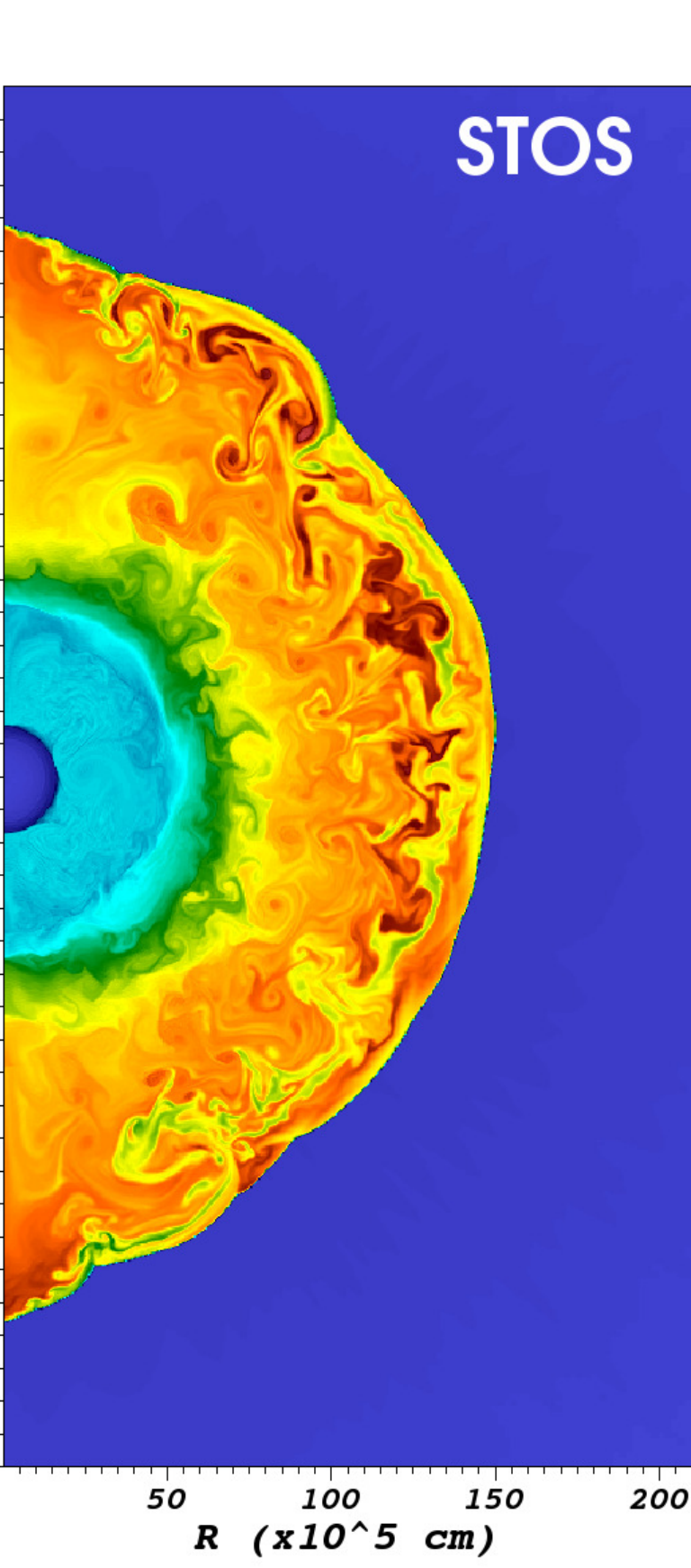}
\end{tabular}
\caption{Entropy for the 2D simulations with $L_{\nu_e,52}=1.3$ at 100 ms post-bounce.  The three different EOS models are shown, from left to right: LS180, LS220, and STOS.}
\label{fig:l13_100ms}
\end{figure*}

\begin{figure*}
\centering
\begin{tabular}{ccc}
\includegraphics[width=2.5in,trim= 0.4in 0.7in 3.4in 0in,clip]{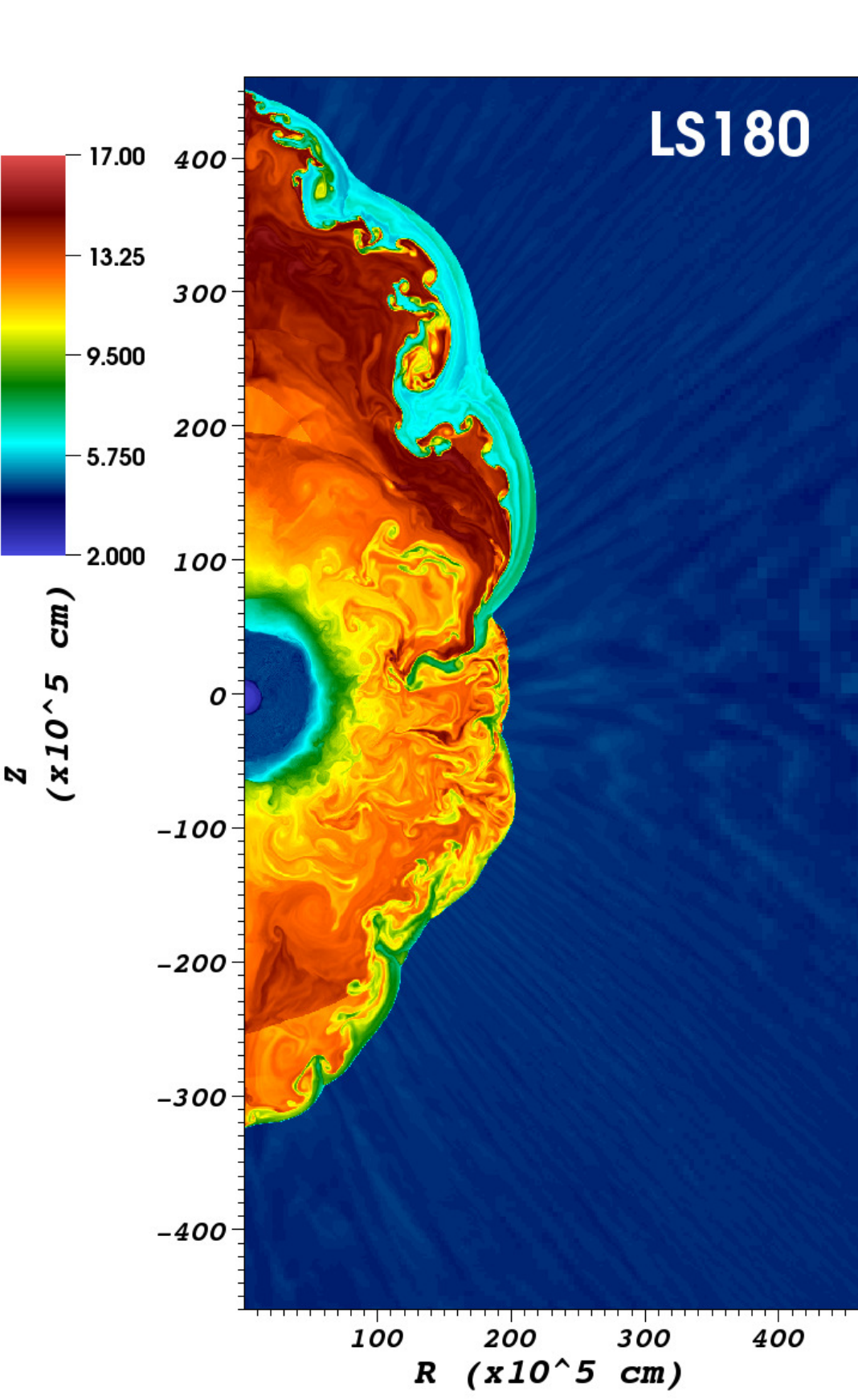}
\includegraphics[width=1.8in,trim= 1.6in 0.7in 3.4in 0.in,clip]{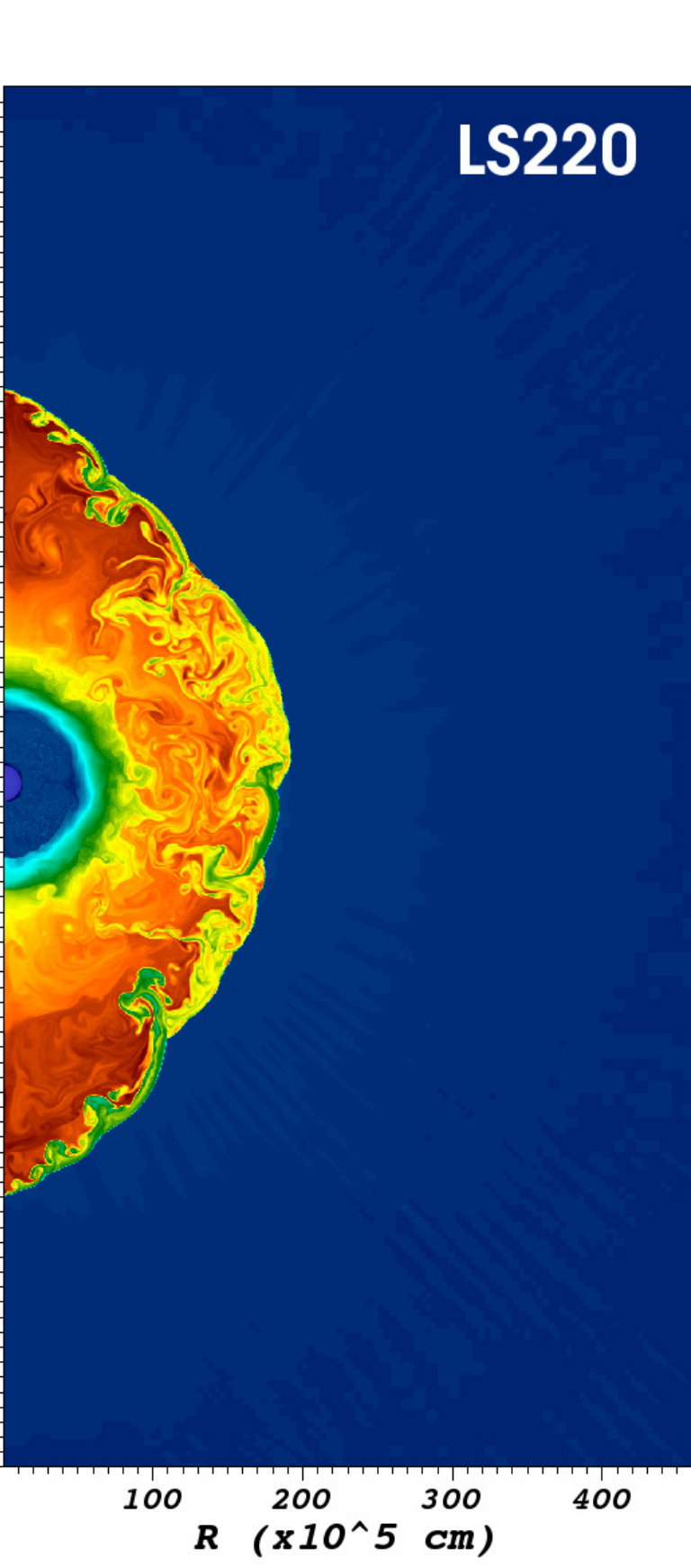}
\includegraphics[width=1.8in,trim= 1.6in 0.7in 3.4in 0in,clip]{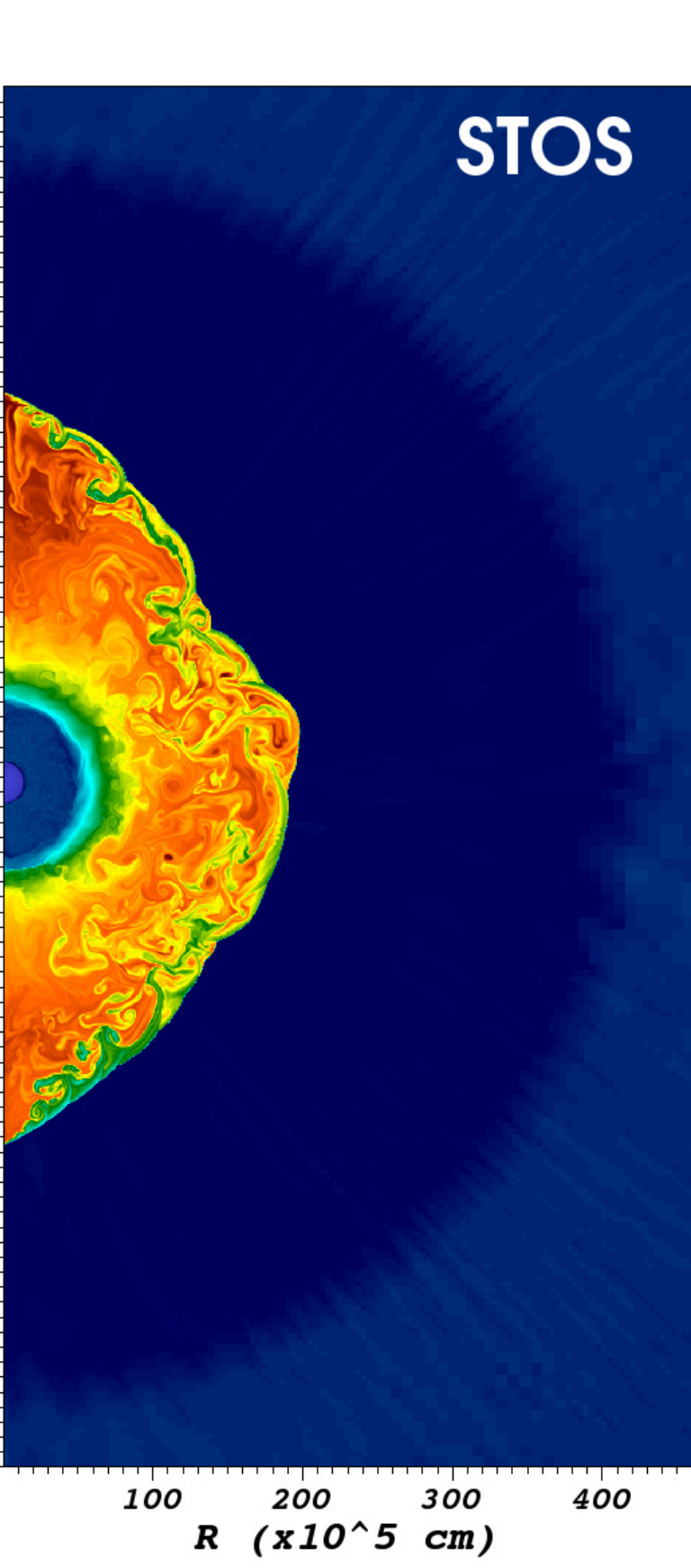}
\end{tabular}
\caption{Same as Fig. \ref{fig:l13_100ms} except at 300 ms post-bounce.}
\label{fig:l13_300ms}
\end{figure*}

\begin{figure*}
\centering
\begin{tabular}{ccc}
\includegraphics[width=2.5in,trim= 0.4in 0.7in 3.4in 0in,clip]{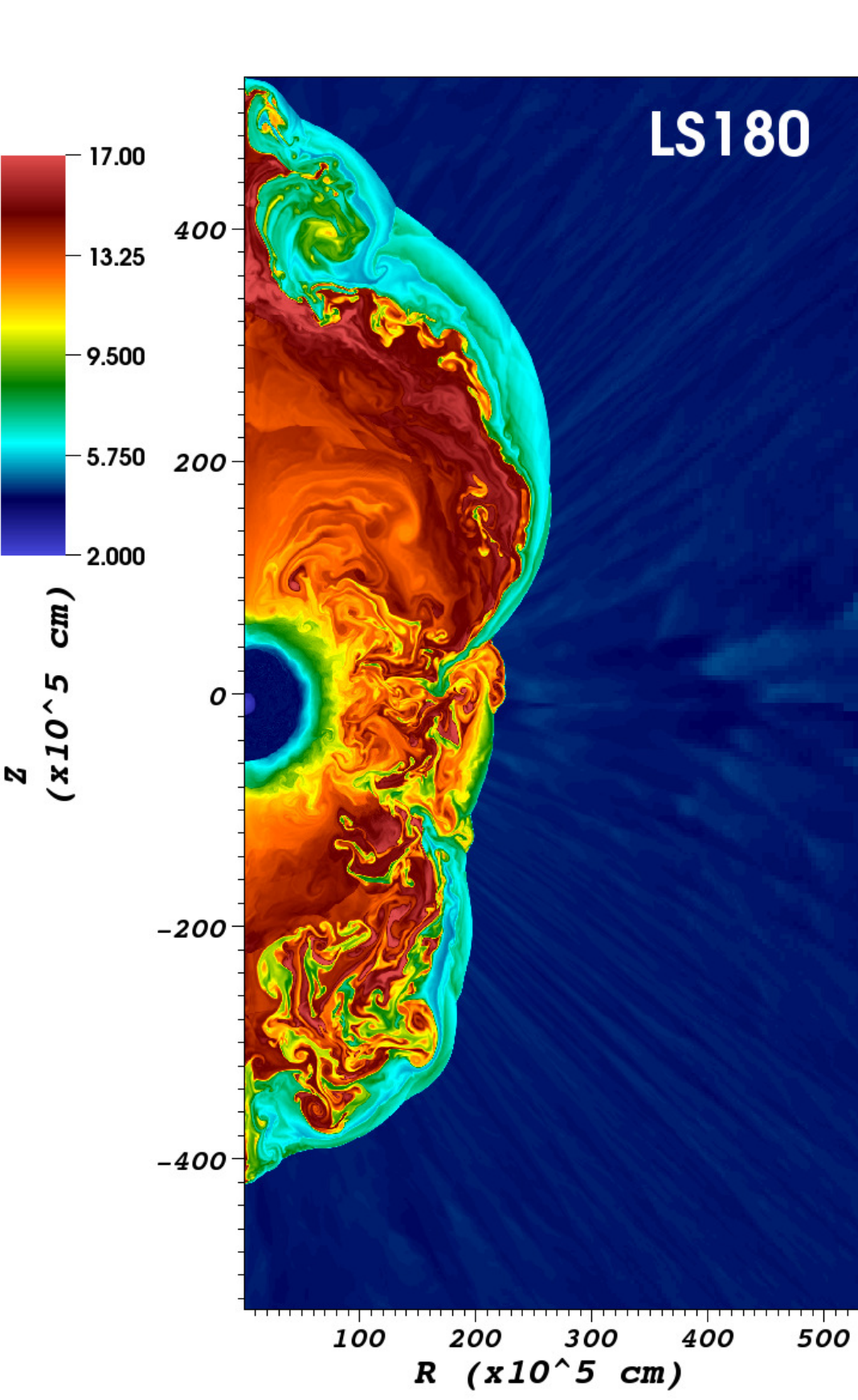}
\includegraphics[width=1.8in,trim= 1.6in 0.7in 3.4in 0.in,clip]{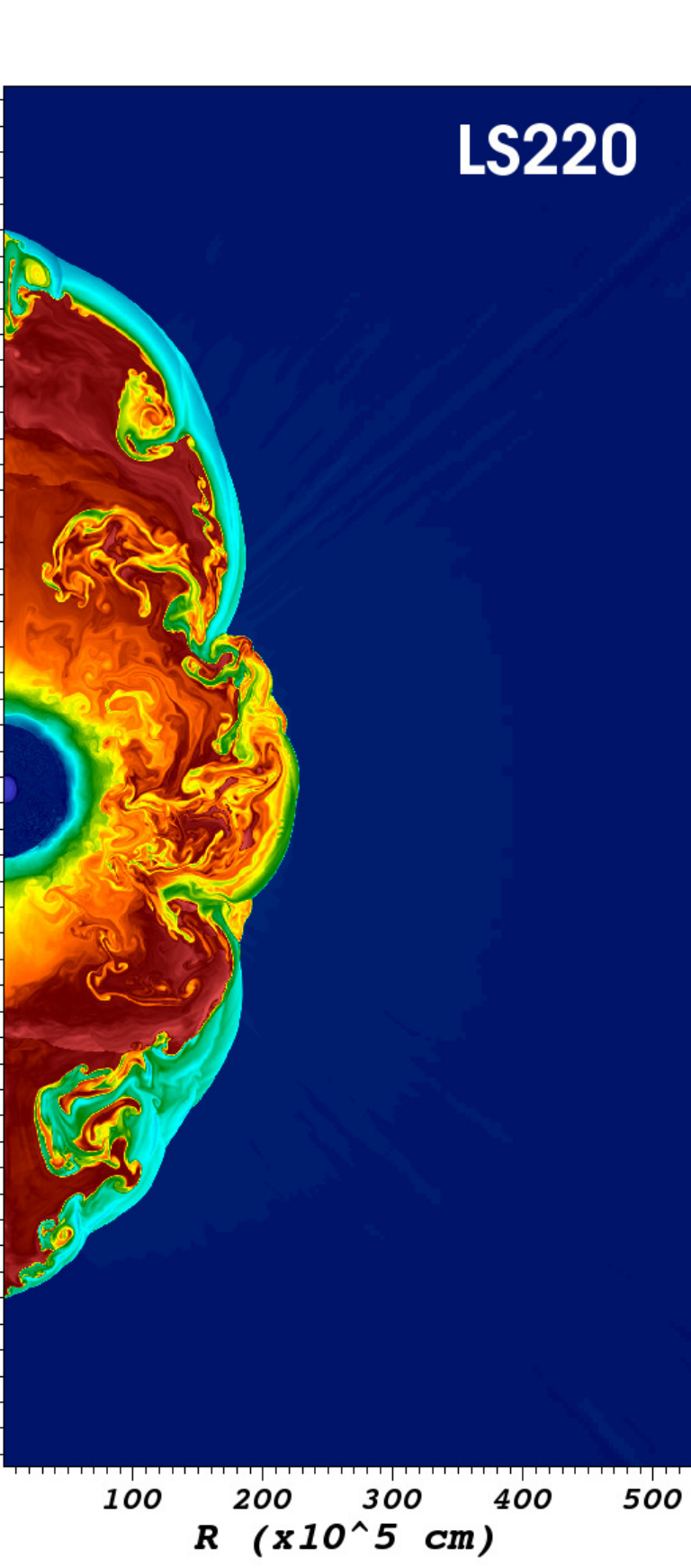}
\includegraphics[width=1.8in,trim= 1.6in 0.7in 3.4in 0in,clip]{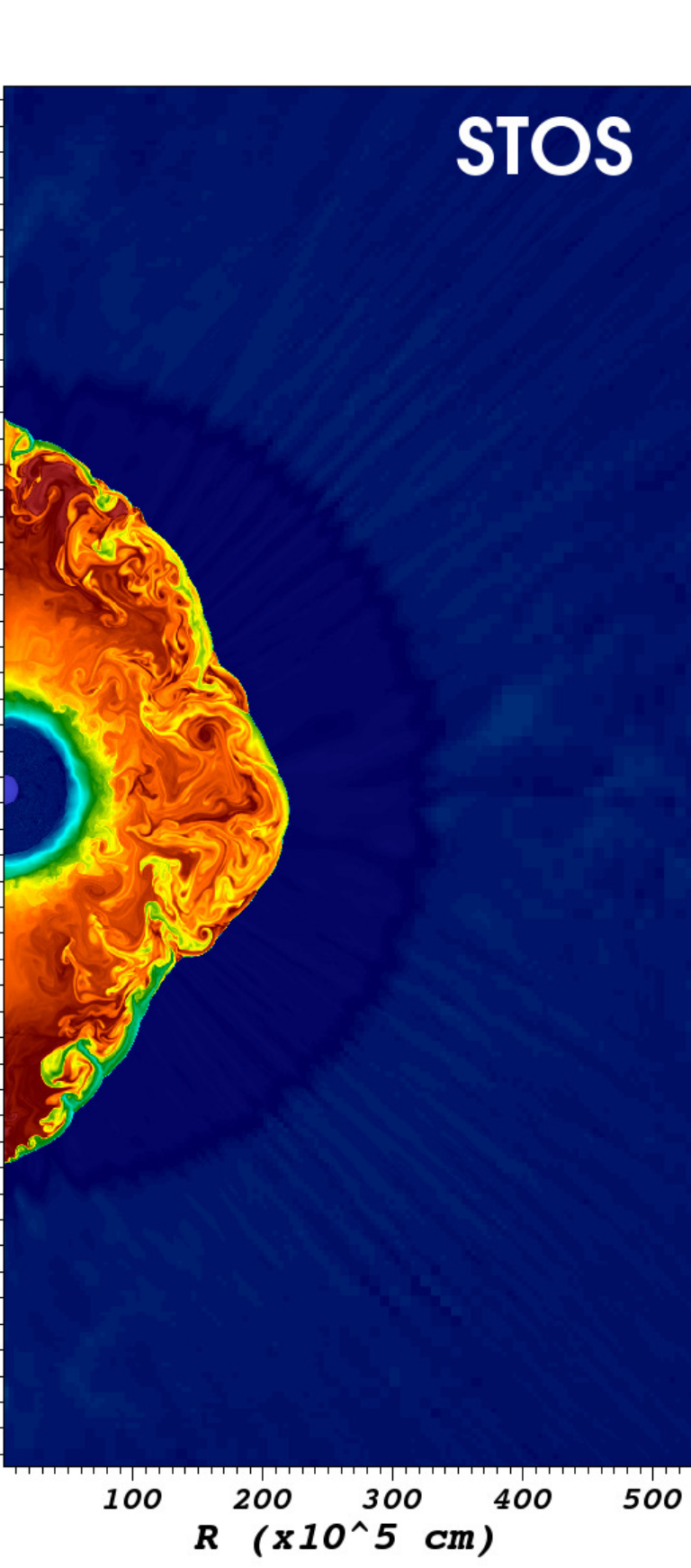}
\end{tabular}
\caption{Same as Fig. \ref{fig:l13_100ms} except at 600 ms post-bounce.}
\label{fig:l13_600ms}
\end{figure*}

Figures \ref{fig:l13_100ms} - \ref{fig:l13_600ms} show the evolution of the entropy in 2D for $L_{\nu_e} = 1.3\times10^{52}\ {\rm erg\ s^{-1}}$.  For this luminosity, only LS180 results in an explosion within 1 second post-bounce.  Figure \ref{fig:l13_100ms} corresponds to 100 ms post-bounce.  At this stage convection has set in in the post-shock region as well as the PNS cooling region (faint blue area in the figure).  Some extension along the axis is evident, particularly for LS180, but a strong $\ell=1$ SASI is not yet apparent.  By 300 ms (Figure \ref{fig:l13_300ms}), however, the SASI is very obvious and is reaching large amplitude for LS180, though less so for LS220 and STOS.  At 600 ms (Figure \ref{fig:l13_600ms}) large-amplitude $\ell=1$ SASI motions are clear for LS180 and LS220, but the SASI amplitude remains modest for STOS.  While LS180 and LS220 are showing shock expansion, particularly along the axis, STOS shows a fairly stationary shock, in an average sense, at around 200 km.

\begin{figure}
\centering
\includegraphics[width=3.5in]{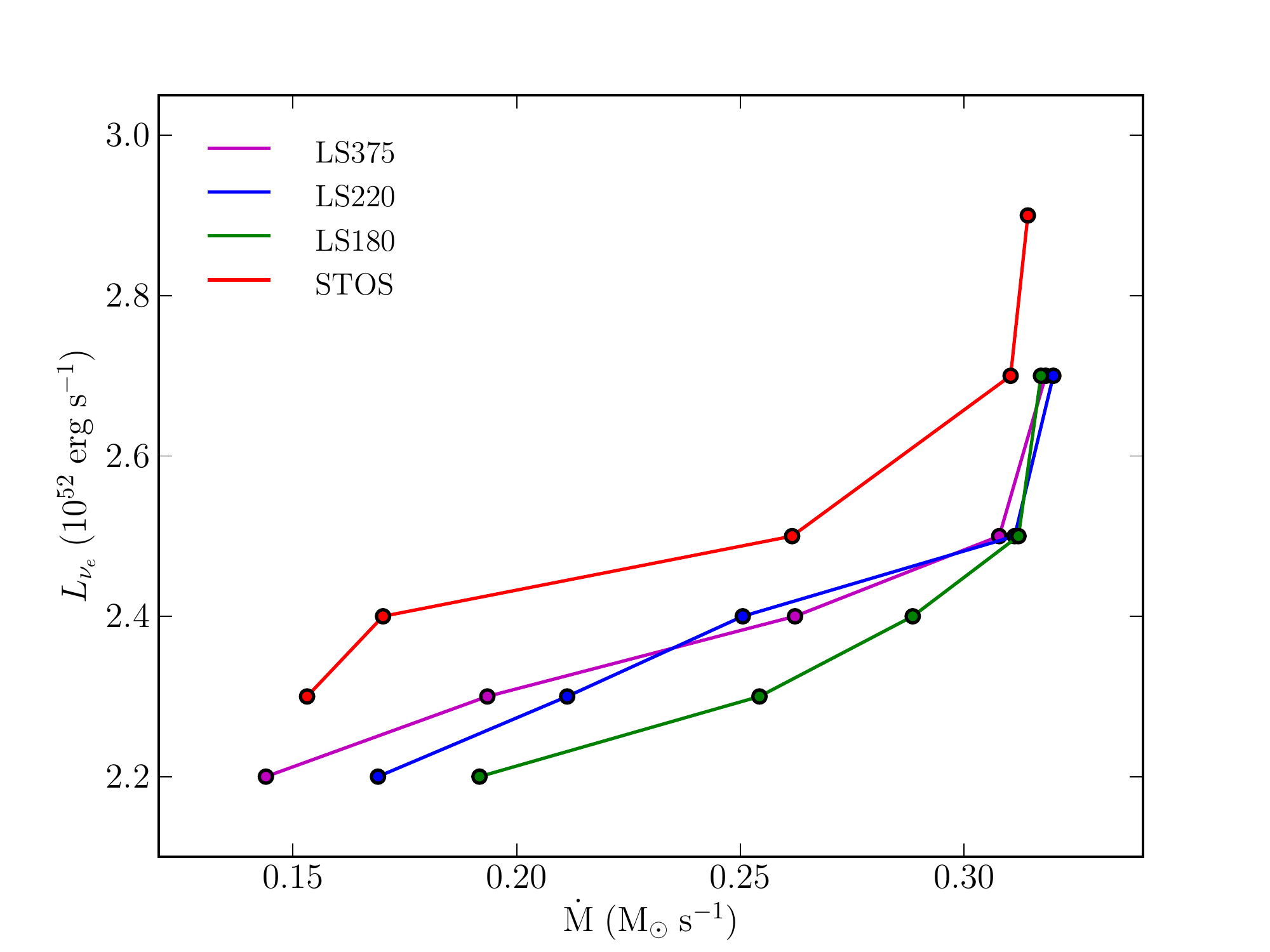}
\caption{The neutrino luminosity versus mass accretion rate at time of explosion for 1D simulations using four different EOS, including LS375.}
\label{fig:ls375}
\end{figure}

Our results show a clear trend with the incompressibility, $K$, of the EOS:  higher values of $K$ result in later explosions.  In order to test the importance of $K$ in determining the time to explosion, we have also run a series of 1D simulations using the Lattimer \& Swesty EOS with $K=375$ MeV.  The results are shown in Figure \ref{fig:ls375} where we plot the neutrino luminosity versus mass accretion rate at time of explosion for our 1D simulations.  As can be seen, the trend with $K$ does not survive: the simulations with LS375 explode earlier (at higher $\dot M$) than STOS.  At low-luminosities, near critical, LS375 explodes later than LS220 and LS180.  At higher luminosities, LS375 roughly converges with LS220, and all three versions of the LS EOS converge at $L_{\nu_e,52} = 2.5$.  This indicates that the time to explosion for a given EOS is not dependent on $K$ alone, but must also depend on the higher-order EOS parameters listed in Table \ref{table:eosParams}.

Why, given an identical neutrino luminosity, do differences in the baryonic EOS result in different evolution of the supernova shock?  That is, what mechanism couples the behavior of the PNS, where the differences between the EOS are important, to the motion of the shock?  The difference in time-to-explosion we find for the various EOS is likely due to EOS-dependent behavior of the SASI.  Whether the underlying SASI mechanism is solely acoustic \citep{Blondin:2006dv} or advective-acoustic \citep{Foglizzo:2007cq} the magnitude of the acoustic energy flux from the PNS is a determining factor in its growth \citep{2012MNRAS.421..546G}. \footnote{\citet{2012MNRAS.421..546G} present a compelling argument that the linear growth of the SASI must be dominated by the advective-acoustic cycle rather than the purely acoustic cycle.}  A soft EOS, such as LS180 and LS220, results in a more compact, higher-density PNS as compared to a stiff EOS like STOS.  This is true for the central core of the PNS as well as the `deceleration' region at the edge of the PNS where advected perturbations are turned into sound waves.  The EOS dependence we find can likely be explained by a difference in the acoustic flux generated at the edge of the PNS:  the denser PNS's produced by LS180 and LS220 generate a larger outbound acoustic flux than that of STOS and, thus, lead to larger amplitude excitations of the SASI.  Dense PNS are, in effect, better sonic transducers.  This will hold in both 2D and 1D, where the SASI is purely radial.  Differences in the growth of the SASI also accounts for the similarity in time-to-explosion for LS180 and LS220 at high $L_{\nu_e}$: these prompt explosions do not depend so much on the SASI.

The higher mass accretion rate of STOS at early times (Figure \ref{fig:massAcc}) is causal to the differences in time-to-explosion for high-$L_{\nu_e}$ simulations.  This higher accretion rate prior to 100 ms post-bounce results in a greater mass of iron being dissociated at the shock and, therefore, more energy loss.   In the high-$L_{\nu_e}$ models, the shock begins running away to large radii at the moment the mass accretion rate drops precipitously (see Figure \ref{fig:rshock}), corresponding to the end of the accretion of the iron core.  Since this is delayed slightly for STOS, the shock begins its runaway expansion later as compared with LS180 and LS220.  This difference will be less important at later times and cannot explain the differences in time-to-explosion between LS180 and LS220 (and LS375 in 1D) we find for lower-$L_{\nu_e}$ models.

\begin{figure*}[!htb]
\centering
\begin{tabular}{cc}
\includegraphics[width=3.25in,trim= 0.2in 0in 0.5in 0.5in,clip]{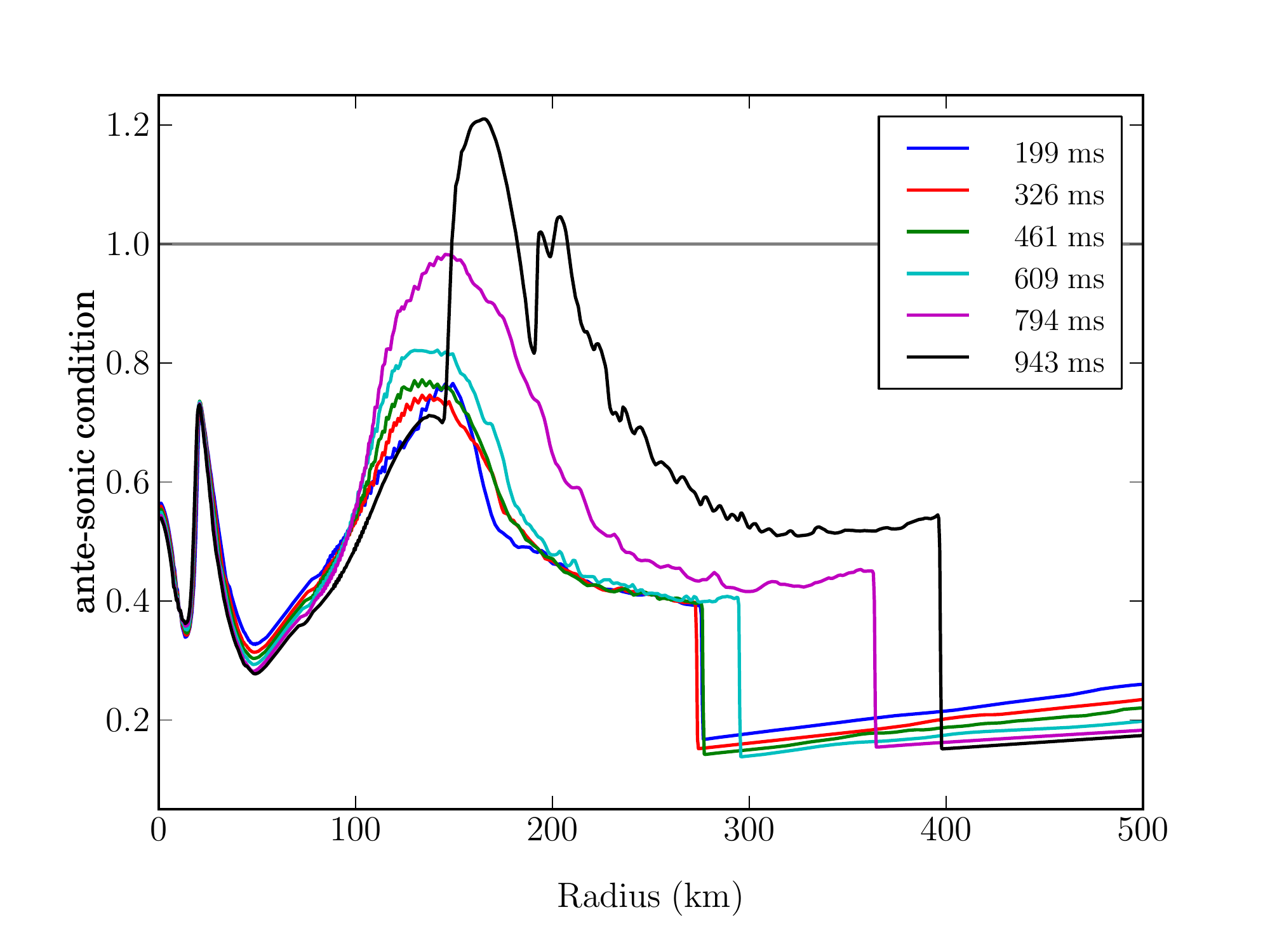}
\includegraphics[width=3.25in,trim= 0.2in 0in 0.5in 0.5in,clip]{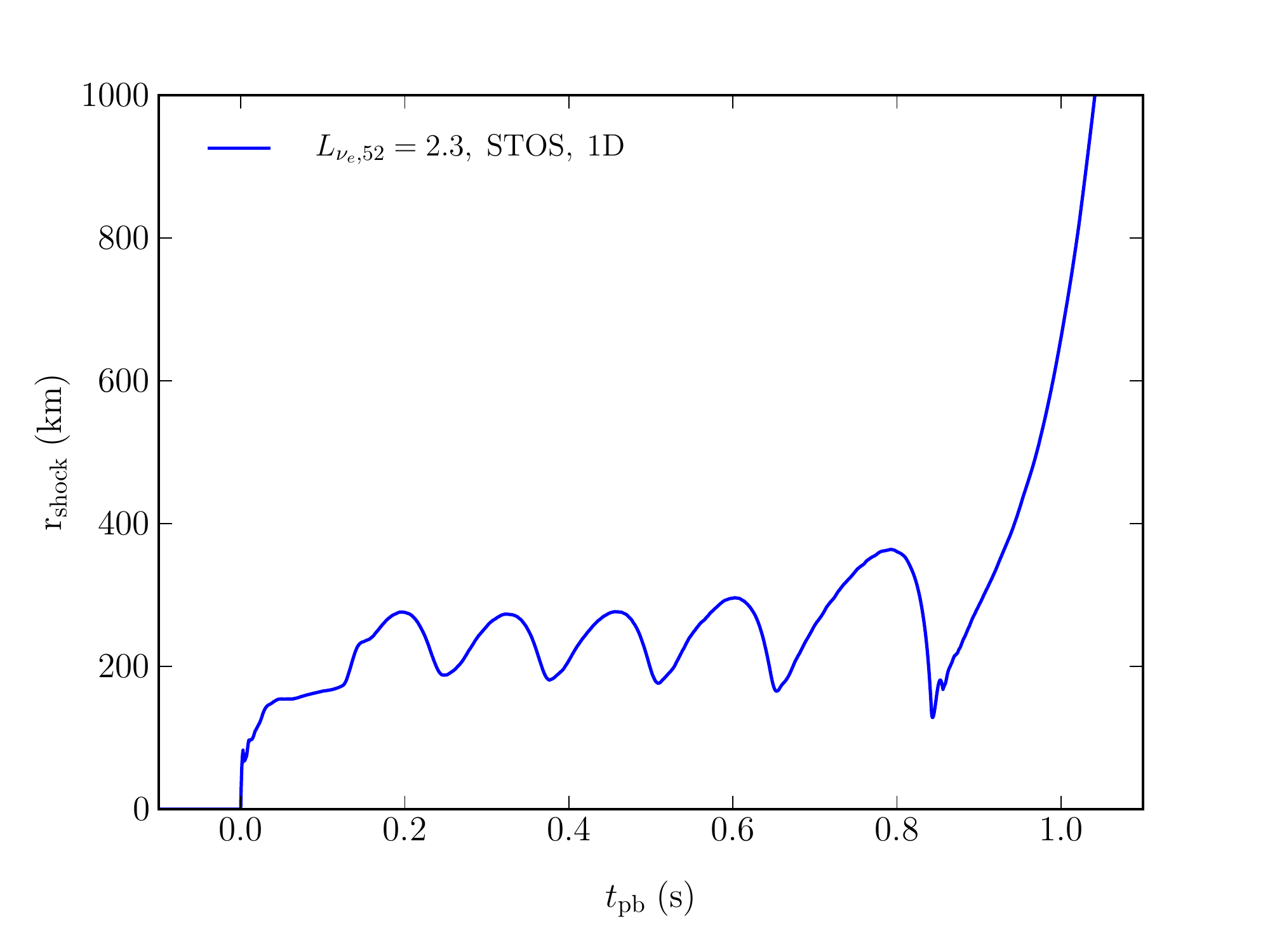}
\end{tabular}
\caption{The ante-sonic condition as a function of radius at several times post-bounce for the STOS model with $L_{\nu_e}=2.3\times10^{52}\ {\rm erg\ s^{-1}}$ (left panel).  Shown for comparison is the shock radius as a function of time for this model (right panel).  The times plotted on the left correspond to maximum shock extensions seen on the right.  Here, the ante-sonic condition is computed as $(c_S^2/v_{\rm esc}^2)/(0.18 \gamma_c)$.  Instability should occur when this quantity exceeds approximately one (gray line in left panel).  As can be seen in this figure, this occurs only during the final expansion of the shock (black line).  }
\label{fig:antes}
\end{figure*}

\subsection{Criterion for Transition to Runaway Shock Expansion}

Recently, \citet{Pejcha:2012cw} and \citet{Fernandez:2012kg} have studied the conditions that lead to explosions in simple, parameterized models of stalled shocks in the CCSN context.  Using principles of stellar pulsation theory, \citet{Fernandez:2012kg} develops an instability criterion that closely resembles the critical condition found in previous simulations:  a greater advection time through the gain region than through the cooling region.  \citet{Pejcha:2012cw} find a criterion based on the ratio of the sound speed to the escape velocity, the so-called `ante-sonic condition.'  We find that both criteria can be good predictors for the transition to runaway shock expansion in our 1D simulations.  The ante-sonic condition is, however, easier to compute (especially in 2D) since it does not involve spatial integrals.  The ante-sonic condition for a polytropic EOS is 
\begin{equation}
{\rm max} \left ( \frac{c_S^2}{v_{\rm esc}^2} \right ) = 0.18 \Gamma,
\end{equation}
where $v_{\rm esc}$ is the escape velocity, and $\Gamma$ is the polytropic index.  In Figure \ref{fig:antes} we plot the quantity $(c_S^2 / v_{\rm esc}^2 )/ (0.18 \gamma_c)$ as a function of space and time for the 1D simulation using the STOS EOS with $L_{\nu_e,52} = 2.3$, along with the shock radius as a function of time for this model.  For the general EOS, we substitute the spatially-varying adiabatic index as returned by the EOS, $\gamma_c$, for the polytropic index in the ante-sonic condition.  In Figure \ref{fig:antes}, instability of the type described by \citet{Pejcha:2012cw} occurs for values greater than one.  The highest values for the ante-sonic relation occur when the shock reaches radial maxima, thus we plot the times when the shock has just started to collapse back during its radial oscillations.  As shown, the ante-sonic condition is only satisfied on the last, explosive expansion of the shock.  This is true for all of our 1D models except the $L_{\nu_e,52}=2.4$ model using STOS; in this model, we see the shock expand well beyond 400 km and then fall back inward one last time before runaway shock expansion occurs (see Figure \ref{fig:rshock}).  During the shock's first expansion beyond 400 km in this model, the ante-sonic condition is met but the shock nevertheless contracts again.

In 2D, we find that the ante-sonic condition is generally met in post-shock regions that have expanded to large radii.  This can happen long before the average shock radius exceeds 400 km when we consider the model to have successfully `exploded.'  For radial rays along which the ante-sonic condition has been met at some point, the shock radius tends to remain above 400 km.

\subsection{Dependence on Resolution}

\begin{figure}
\centering
\includegraphics[width=3.25in,trim= 0.2in 0in 0.5in 0.5in,clip]{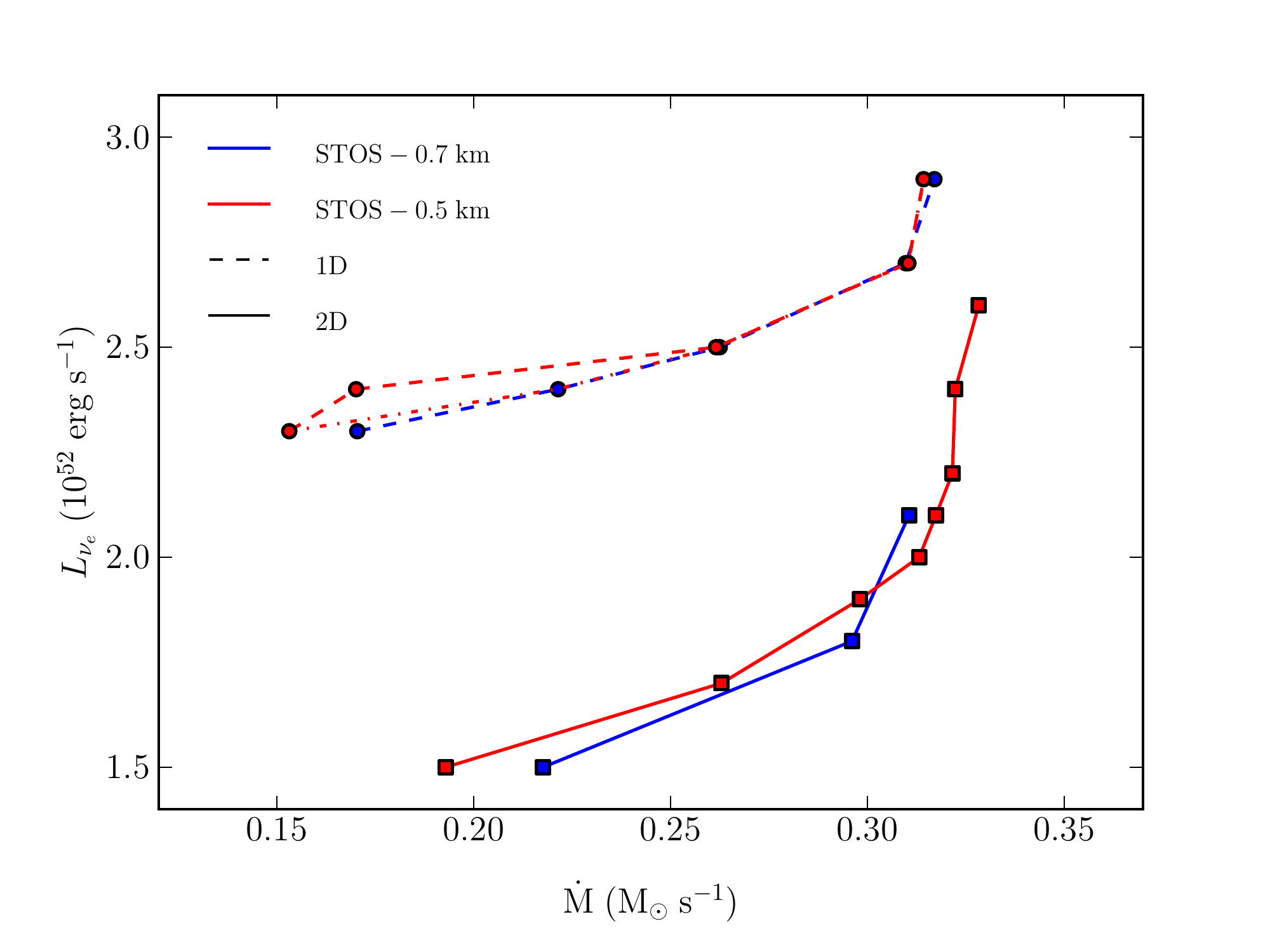}
\caption{Neutrino luminosity versus mass accretion rate for 1D and 2D simulations with STOS at two different resolutions: the fiducial 0.5 km and 0.7 km.  For 1D, the dash-dot curve represents using the first time the shock crosses 400 km as the explosion time for the $L_{\nu_e,52}=2.4$ model.  Doing this makes the curves at different resolutions essentially identical.  In both the 1D and 2D cases, the resulting curves shown here are very similar, indicating our results are roughly converged in spatial resolution.}
\label{fig:resolution}
\end{figure}

\citet{Hanke:2011vc} report that their simulations, which employ the same parametric neutrino heating/cooling as this study, are significantly resolution-dependent.  In their 2D simulations, finer resolution led to earlier explosions for the same driving neutrino luminosity.  We have examined the dependence of both our 1D and 2D results on resolution and show the results in Figure \ref{fig:resolution}.  While the development of small-scale instabilities and turbulence will depend heavily on the resolution used, for the present study the important question is whether or not changing the resolution significantly alter the time at which a model explodes for a given neutrino luminosity.  To determine this, we ran several models using STOS with an increased minimum grid spacing at the maximum level of refinement of 0.7 km, versus 0.5 km for our fiducial models.  We find the results for explosion times as a function of neutrino luminosity are essentially identical in 1D and not significantly different in 2D.  As shown in Figure \ref{fig:resolution}, the 1D curves lie on top of one another and the 2D curves are very similar.  For 2D, the small differences are that at lower driving luminosities (and, hence, later explosion times) the low-resolution models explode earlier, while at higher luminosities the lower resolution models explode later.  Thus the high-resolution and low-resolution curves cross each other in Figure \ref{fig:resolution}.

It is difficult to make a direct comparison between our results and those of \citet{Hanke:2011vc} because the nature of the computational grids employed is quite different.  \citet{Hanke:2011vc} use 2D spherical-polar coordinates with non-equidistant spacing in radius.  Here we use 2D cylindrical $R-z$ coordinates with adaptive mesh refinement.\footnote{\citet{Nordhaus:2010ct} also use 2D cylindrical coordinates with AMR.}  Hanke et al. use a fiducial resolution of $3^\circ$ in angle and 400 zones in radius.  At a radius of 100 km, near the base of the gain region, this yields a linear grid spacing in the angular direction of 1.67 km, more than three times our fiducial resolution.  For a spherical grid, this grid spacing will increase with radius while in our simulations we allow refinement to the highest level anywhere in the computational domain, allowing us to achieve 0.5 km resolution for the entire post-shock region at all times.  The highest resolution simulations of Hanke et al. use 0.5$^\circ$ angular resolution, resulting in 0.28 km linear resolution at 100 km and 1.1 km linear resolution at 400 km.  This resolution is likely comparable to our fiducial resolution, the linear grid spacing being smaller at the base of the gain region and larger at radii where shock oscillations reach the greatest extent.

We should not expect hydrodynamic simulations that involve instabilities and turbulence, such as these, to converge in spatial resolution, but the fact that we do not see significant differences in the explosion times at different resolutions indicates that our results are converged with respect to the result we seek.  

\section{Conclusions}
\label{sec:Conclusions}

We have conducted a parameter study exploring the dependence of the neutrino mechanism of core-collapse supernovae on the equation of state.  We varied the driving neutrino luminosity and EOS in several 1D and 2D simulations of stellar core collapse and measured the resulting explosion times and mass accretion rates at the time of explosion.  Table \ref{table:results} and Figure \ref{fig:massAcc} summarize our results.  We find that for every EOS, 2D explosion are obtained more easily than in 1D, and that  models using the Lattimer \& Swesty EOS explodes more easily than models using the EOS of Shen et al., with LS180 resulting in slightly earlier explosions than LS220.  Thus, our results show a general trend that softer EOS lead to easier explosions.  In order to test if the incompressibility parameter, $K$, is the primary determinant of the time delay to explosion for a given neutrino luminosity, we also ran a series of 1D simulations using LS375.  This model has a substantially higher incompressibility than STOS (375 MeV v. 281 MeV).  We find that STOS, despite have a lower incompressibility, explodes later than LS375 for all neutrino luminosities.  This highlights that the higher-order EOS parameters, such as symmetry energy coefficient, play an important role in determining the stiffness of an EOS.  Our results of explosion delay times do follow a trend in resulting neutron star radii (see Fig. \ref{fig:TOV}).  EOS that produce larger typical neutron star radii explode later in our simulations.  The dependence of the time-to-explosion on the EOS is likely due to the dependence of the acoustic energy flux from the PNS on the stiffness of the EOS.   The softer EOS of Lattimer \& Swesty result in more compact PNS that are more effective at translating advected perturbations into outgoing sound waves.  This then results in more acoustic excitation of the standing accretion shock instability and an overall faster rate of shock expansion.  This should be true no matter if the underlying SASI mechanism is acoustic \citep{Blondin:2006dv} or advective-acoustic \citep{Foglizzo:2007cq} since both mechanisms depend on the acoustic flux from the PNS.  For high-$L_{\nu_e}$ models that do not depend strongly on the SASI, we attribute the difference between STOS and LS to the higher mass accretion rate just after bounce for STOS.

In this study, we use a parameterized neutrino ``light-bulb" with a fixed neutrino luminosity.  There is no back-reaction from the EOS on the neutrino luminosity, allowing us to examine the dependence of the resulting explosion times on the EOS alone.  Previous studies have shown that the neutrino luminosity can be significantly dependent on the EOS \citep{Thompson:2003kn, Sumiyoshi:2005kg} and this is a very important consequence of the choice of EOS, but our study shows that even for a fixed neutrino luminosity, the likelihood of a given model to explode also depends on the EOS.  

The EOS for matter at temperatures and densities relevant to CCSNe is still an active area of research.  As summarized in the Introduction and Table \ref{table:eosParams}, available experiments and observations that constrain the EOS parameters favor LS220.  LS180 is ruled out as it cannot produce a $\sim 2 M_\sun$ neutron star as required by the observations of PSR J1614-2230 \citep{Demorest:2010bf}.  STOS yields neutron stars of typical mass (around 1.4$M_\sun$) with too large a radius ($\sim15$ km) to account for the results of \citet{Steiner:2010el} who estimate typical NS radii of 11-12 km.  Several of the EOS parameters listed in Table \ref{table:eosParams} for STOS also fall outside of the experimental and theoretical estimates.  So, LS220 is the EOS that satisfies, or nearly satisfies, the largest number of constraints.  Our results show, additionally, that LS220 is intermediate between STOS and LS180 in favoring explosion for a given neutrino luminosity.  Until we are more certain about the EOS at temperatures and densities relevant to CCSN simulations, LS220 seems a good choice.  Some recent works have chosen to use LS220 on these same grounds \citep{Ott:2012ta, Dessart:2012vi}.

\acknowledgements
We are grateful to Craig Wheeler, Christian Ott, Milo\v s Milosavljevi\'c, and Dongwook Lee for helpful discussions.  We thank Evan O'Connor for providing help with implementing the EOS routines used in this work, for providing EOS data, and for very insightful discussions.  Support for this work was provided by NASA through Hubble Fellowship grant No. 51286.01 awarded by the Space Telescope Science Institute, which is operated by the Association of Universities for Research in Astronomy, Inc., for NASA, under contract NAS 5-26555.  The software used in this work was in part developed by the DOE NNSA-ASC OASCR Flash Center at the University of Chicago.  This research used computational resources at ALCF at ANL, which is supported by the Office of Science of the US Department of Energy under Contract No. DE-AC02-06CH11357.  The author acknowledges the Texas Advanced Computing Center (TACC) at The University of Texas at Austin for providing high-performance computing, visualization, and data storage resources that have contributed to the research results reported within this paper

\bibliography{References}

\begin{thebibliography}{62}
\expandafter\ifx\csname natexlab\endcsname\relax\def\natexlab#1{#1}\fi

\bibitem[{Balsara(2004)}]{Balsara:2004cz}
Balsara, D.~S. 2004, \apjs, 151, 149

\bibitem[{Baron {et~al.}(1985{\natexlab{a}})Baron, Cooperstein, \&
  Kahana}]{Baron:1985ec}
Baron, E., Cooperstein, J., \& Kahana, S. 1985{\natexlab{a}}, Nuclear Physics
  A, 440, 744

\bibitem[{Baron {et~al.}(1985{\natexlab{b}})Baron, Cooperstein, \&
  Kahana}]{Baron:1985cd}
---. 1985{\natexlab{b}}, Phys. Rev. Lett., 55, 126

\bibitem[{Bethe \& Wilson(1985)}]{Bethe:1985da}
Bethe, H.~A., \& Wilson, J.~R. 1985, \apj, 295, 14

\bibitem[{{Blondin} \& {Mezzacappa}(2006)}]{Blondin:2006dv}
{Blondin}, J.~M., \& {Mezzacappa}, A. 2006, \apj, 642, 401

\bibitem[{Blondin {et~al.}(2003)Blondin, Mezzacappa, \&
  DeMarino}]{Blondin:2003ep}
Blondin, J.~M., Mezzacappa, A., \& DeMarino, C. 2003, \apj, 584, 971

\bibitem[{Bruenn {et~al.}(2009)Bruenn, Mezzacappa, Hix, Blondin, Marronetti,
  Messer, Dirk, \& Yoshida}]{Bruenn:2009cw}
Bruenn, S.~W., Mezzacappa, A., Hix, W.~R., Blondin, J.~M., Marronetti, P.,
  Messer, O. E.~B., Dirk, C.~J., \& Yoshida, S. 2009, Journal of Physics:
  Conference Series, 180, 2018

\bibitem[{Burrows {et~al.}(2007{\natexlab{a}})Burrows, Dessart, Ott, \&
  Livne}]{Burrows:2007kha}
Burrows, A., Dessart, L., Ott, C.~D., \& Livne, E. 2007{\natexlab{a}}, Physics
  Reports, 442, 23

\bibitem[{Burrows \& Fryxell(1993)}]{Burrows:1993ki}
Burrows, A., \& Fryxell, B.~A. 1993, \apjl, 418, L33

\bibitem[{Burrows {et~al.}(2006)Burrows, Livne, Dessart, Ott, \&
  Murphy}]{Burrows:2006js}
Burrows, A., Livne, E., Dessart, L., Ott, C.~D., \& Murphy, J. 2006, \apj, 640,
  878

\bibitem[{Burrows {et~al.}(2007{\natexlab{b}})Burrows, Livne, Dessart, Ott, \&
  Murphy}]{Burrows:2007kh}
---. 2007{\natexlab{b}}, \apj, 655, 416

\bibitem[{Colella \& Woodward(1984)}]{1984JCoPh..54..174C}
Colella, P., \& Woodward, P.~R. 1984, Journal of Computational Physics (ISSN
  0021-9991), 54, 174

\bibitem[{Colgate \& White(1966)}]{Colgate:1966cl}
Colgate, S.~A., \& White, R.~H. 1966, \apj, 143, 626

\bibitem[{Demorest {et~al.}(2010)Demorest, Pennucci, Ransom, Roberts, \&
  Hessels}]{Demorest:2010bf}
Demorest, P.~B., Pennucci, T., Ransom, S.~M., Roberts, M. S.~E., \& Hessels, J.
  W.~T. 2010, Nature, 467, 1081

\bibitem[{Dessart {et~al.}(2006)Dessart, Burrows, Livne, \&
  Ott}]{Dessart:2006cg}
Dessart, L., Burrows, A., Livne, E., \& Ott, C.~D. 2006, \apj, 645, 534

\bibitem[{Dessart {et~al.}(2012)Dessart, O'Connor, \& Ott}]{Dessart:2012vi}
Dessart, L., O'Connor, E., \& Ott, C.~D. 2012, ArXiv e-prints

\bibitem[{Dubey {et~al.}(2009)Dubey, Antypas, Ganapathy, \&
  Reid}]{Dubey:2009hh}
Dubey, A., Antypas, K., Ganapathy, M., \& Reid, L. 2009, Parallel Computing

\bibitem[{Epstein(1979)}]{Epstein:1979tg}
Epstein, R.~I. 1979, \mnras, 188, 305

\bibitem[{Fern{\'a}ndez(2012)}]{Fernandez:2012kg}
Fern{\'a}ndez, R. 2012, \apj, 749, 142

\bibitem[{Fischer {et~al.}(2009)Fischer, Whitehouse, Mezzacappa, Thielemann, \&
  Liebend{\"o}rfer}]{Fischer:2009ka}
Fischer, T., Whitehouse, S.~C., Mezzacappa, A., Thielemann, F.-K., \&
  Liebend{\"o}rfer, M. 2009, \aap, 499, 1

\bibitem[{Foglizzo {et~al.}(2007)Foglizzo, Galletti, Scheck, \&
  Janka}]{Foglizzo:2007cq}
Foglizzo, T., Galletti, P., Scheck, L., \& Janka, H.-T. 2007, \apj, 654, 1006

\bibitem[{Furusawa {et~al.}(2011)Furusawa, Yamada, Sumiyoshi, \&
  Suzuki}]{Furusawa:2011ck}
Furusawa, S., Yamada, S., Sumiyoshi, K., \& Suzuki, H. 2011, \apj, 738, 178

\bibitem[{{Guilet} \& {Foglizzo}(2012)}]{2012MNRAS.421..546G}
{Guilet}, J., \& {Foglizzo}, T. 2012, \mnras, 421, 546

\bibitem[{Hanke {et~al.}(2011)Hanke, Marek, Mueller, \& Janka}]{Hanke:2011vc}
Hanke, F., Marek, A., Mueller, B., \& Janka, H.-T. 2011, arXiv, astro-ph.SR

\bibitem[{Hebeler {et~al.}(2010)Hebeler, Lattimer, Pethick, \&
  Schwenk}]{Hebeler:2010dx}
Hebeler, K., Lattimer, J.~M., Pethick, C.~J., \& Schwenk, A. 2010, Physical
  Review Letters, 105, 161102

\bibitem[{Hempel {et~al.}(2012)Hempel, Fischer, Schaffner-Bielich, \&
  Liebend{\"o}rfer}]{Hempel:2012bh}
Hempel, M., Fischer, T., Schaffner-Bielich, J., \& Liebend{\"o}rfer, M. 2012,
  \apj, 748, 70

\bibitem[{Hempel \& Schaffner-Bielich(2010)}]{Hempel:2010fh}
Hempel, M., \& Schaffner-Bielich, J. 2010, Nuclear Physics A, 837, 210

\bibitem[{Hillebrandt \& Wolff(1985)}]{Hillebrandt:1985to}
Hillebrandt, W., \& Wolff, R.~G. 1985, in Nucleosynthesis : Challenges and New
  Developments. Edited by W. David Arnett and James W. Truran. Chicago :
  University of Chicago Press, c1985., p.131, 131

\bibitem[{Janka(2001)}]{Janka:2001fp}
Janka, H.-T. 2001, \aap, 368, 527

\bibitem[{Janka \& Mueller(1996)}]{1996A&A...306..167J}
Janka, H.-T., \& Mueller, E. 1996, \aap, 306, 167

\bibitem[{Lattimer \& Swesty(1991)}]{Lattimer:1991fz}
Lattimer, J.~M., \& Swesty, D.~F. 1991, Nuclear Physics A, 535, 331

\bibitem[{Lee \& Deane(2009)}]{Lee:2009kq}
Lee, D., \& Deane, A.~E. 2009, Journal of Computational Physics, 228, 952

\bibitem[{Lentz {et~al.}(2012)Lentz, Mezzacappa, Bronson~Messer,
  Liebend{\"o}rfer, Hix, \& Bruenn}]{Lentz:2012fy}
Lentz, E.~J., Mezzacappa, A., Bronson~Messer, O.~E., Liebend{\"o}rfer, M., Hix,
  W.~R., \& Bruenn, S.~W. 2012, \apj, 747, 73

\bibitem[{Liebend{\"o}rfer(2005)}]{Liebendorfer:2005ft}
Liebend{\"o}rfer, M. 2005, \apj, 633, 1042

\bibitem[{Liebend{\"o}rfer {et~al.}(2009)Liebend{\"o}rfer, Whitehouse, \&
  Fischer}]{Liebendorfer:2009kw}
Liebend{\"o}rfer, M., Whitehouse, S.~C., \& Fischer, T. 2009, \apj, 698, 1174

\bibitem[{MacNeice {et~al.}(2000)MacNeice, Olson, Mobarry, de~Fainchtein, \&
  Packer}]{2000CoPhC.126..330M}
MacNeice, P., Olson, K.~M., Mobarry, C., de~Fainchtein, R., \& Packer, C. 2000,
  Computer Physics Communications, 126, 330

\bibitem[{Marek \& Janka(2009)}]{Marek:2007vi}
Marek, A., \& Janka, H.-T. 2009, \apj, 694, 664

\bibitem[{Marek {et~al.}(2009)Marek, Janka, \& M{\"u}ller}]{Marek:2009bx}
Marek, A., Janka, H.-T., \& M{\"u}ller, E. 2009, \aap, 496, 475

\bibitem[{Mignone {et~al.}(2005)Mignone, Plewa, \& Bodo}]{Mignone:2005hc}
Mignone, A., Plewa, T., \& Bodo, G. 2005, \apjs, 160, 199

\bibitem[{Mueller {et~al.}(2012)Mueller, Janka, \& Marek}]{Mueller:2012tp}
Mueller, B., Janka, H.-T., \& Marek, A. 2012, ArXiv e-prints

\bibitem[{Murphy \& Burrows(2008)}]{Murphy:2008ij}
Murphy, J.~W., \& Burrows, A. 2008, \apj, 688, 1159

\bibitem[{Murphy \& Meakin(2011)}]{Murphy:2011ci}
Murphy, J.~W., \& Meakin, C. 2011, \apj, 742, 74

\bibitem[{Nordhaus {et~al.}(2010)Nordhaus, Burrows, Almgren, \&
  Bell}]{Nordhaus:2010ct}
Nordhaus, J., Burrows, A., Almgren, A., \& Bell, J. 2010, \apj, 720, 694

\bibitem[{O'Connor \& Ott(2010)}]{OConnor:2010bi}
O'Connor, E., \& Ott, C.~D. 2010, Classical and Quantum Gravity, 27, 114103

\bibitem[{O'Connor \& Ott(2011)}]{OConnor:2011hk}
---. 2011, \apj, 730, 70

\bibitem[{Ott {et~al.}(2012)Ott, Abdikamalov, O'Connor, Reisswig, Haas, Kalmus,
  Drasco, Burrows, \& Schnetter}]{Ott:2012ta}
Ott, C.~D., Abdikamalov, E., O'Connor, E., Reisswig, C., Haas, R., Kalmus, P.,
  Drasco, S., Burrows, A., \& Schnetter, E. 2012, ArXiv e-prints

\bibitem[{Pejcha \& Thompson(2012)}]{Pejcha:2012cw}
Pejcha, O., \& Thompson, T.~A. 2012, \apj, 746, 106

\bibitem[{Quirk(1994)}]{Quirk:1994co}
Quirk, J.~J. 1994, International Journal for Numerical Methods in Fluids, 18,
  555

\bibitem[{Shen {et~al.}(2011{\natexlab{a}})Shen, Horowitz, \&
  O'Connor}]{Shen:2011ta}
Shen, G., Horowitz, C., \& O'Connor, E. 2011{\natexlab{a}}, Physical Review C,
  83, 65808

\bibitem[{Shen {et~al.}(2010)Shen, Horowitz, \& Teige}]{Shen:2010wa}
Shen, G., Horowitz, C., \& Teige, S. 2010, Physical Review C, 82, 489

\bibitem[{Shen {et~al.}(2011{\natexlab{b}})Shen, Horowitz, \&
  Teige}]{Shen:2011wk}
---. 2011{\natexlab{b}}, Physical Review C, 83, 35802

\bibitem[{Shen {et~al.}(1998)Shen, Toki, Oyamatsu, \& Sumiyoshi}]{Shen:1998kx}
Shen, H., Toki, H., Oyamatsu, K., \& Sumiyoshi, K. 1998, Nuclear Physics A,
  637, 435

\bibitem[{Shen {et~al.}(2011{\natexlab{c}})Shen, Toki, Oyamatsu, \&
  Sumiyoshi}]{2011ApJS..197...20S}
---. 2011{\natexlab{c}}, \apjs, 197, 20

\bibitem[{Skinner \& Ostriker(2010)}]{Skinner:2010vl}
Skinner, A., \& Ostriker, E. 2010, eprint arXiv, 1004, 2487

\bibitem[{Steiner {et~al.}(2010)Steiner, Lattimer, \& Brown}]{Steiner:2010el}
Steiner, A.~W., Lattimer, J.~M., \& Brown, E.~F. 2010, \apj, 722, 33

\bibitem[{Sumiyoshi {et~al.}(2007)Sumiyoshi, Yamada, \&
  Suzuki}]{Sumiyoshi:2007cd}
Sumiyoshi, K., Yamada, S., \& Suzuki, H. 2007, \apj, 667, 382

\bibitem[{Sumiyoshi {et~al.}(2005)Sumiyoshi, Yamada, Suzuki, Shen, Chiba, \&
  Toki}]{Sumiyoshi:2005kg}
Sumiyoshi, K., Yamada, S., Suzuki, H., Shen, H., Chiba, S., \& Toki, H. 2005,
  \apj, 629, 922

\bibitem[{Suwa {et~al.}(2010)Suwa, Kotake, Takiwaki, Whitehouse,
  Liebend{\"o}rfer, \& Sato}]{Suwa:2010wp}
Suwa, Y., Kotake, K., Takiwaki, T., Whitehouse, S.~C., Liebend{\"o}rfer, M., \&
  Sato, K. 2010, \pasj, 62, L49

\bibitem[{Takiwaki {et~al.}(2012)Takiwaki, Kotake, \& Suwa}]{Takiwaki:2012ck}
Takiwaki, T., Kotake, K., \& Suwa, Y. 2012, \apj, 749, 98

\bibitem[{Thompson {et~al.}(2003)Thompson, Burrows, \& Pinto}]{Thompson:2003kn}
Thompson, T.~A., Burrows, A., \& Pinto, P.~A. 2003, \apj, 592, 434

\bibitem[{Weinberg \& Quataert(2008)}]{Weinberg:2008ky}
Weinberg, N.~N., \& Quataert, E. 2008, \mnras: Letters, 387, L64

\bibitem[{Woosley \& Weaver(1995)}]{Woosley:1995jn}
Woosley, S.~E., \& Weaver, T.~A. 1995, \apjs, 101, 181

\end{thebibliography}

\end{document}